\documentclass[12pt,preprint]{aastex}

\shorttitle{MINOR PLANETS AT WHITE DWARFS}
\shortauthors{FARIHI, JURA, \& ZUCKERMAN}

\begin{document}

\title{INFRARED SIGNATURES OF DISRUPTED\\ MINOR PLANETS AT WHITE DWARFS}

\author{J. Farihi\altaffilmark{1,2,3},
			M. Jura\altaffilmark{2}, and
	 		B. Zuckerman\altaffilmark{2}}

\altaffiltext{1}{Department of Physics \& Astronomy,
			University of Leicester,
			University Road,
			Leicester LE1 7RH, 
			UK; 
			jf123@star.le.ac.uk}

\altaffiltext{2}{Department of Physics \& Astronomy,
			University of California,
			430 Portola Plaza,
			Los Angeles, CA 90095; 
			jura,ben@astro.ucla.edu}
			
\altaffiltext{3}{Gemini Observatory,
			Northern Operations,
			670 North A'ohoku Place,
			Hilo, HI 96720}
			
\begin{abstract}

{\em Spitzer Space Observatory} IRAC and MIPS photometric observations are presented for 
20 white dwarfs with $T_{\rm eff}\la20,000$ K and metal-contaminated photospheres.  A warm 
circumstellar disk is detected at GD 16 and likely at PG 1457$-$086, while the remaining targets 
fail to reveal mid-infrared excess typical of dust disks, including a number of heavily polluted 
stars.  Extending previous studies, over 50\% of all single white dwarfs with implied metal 
accretion rates $dM/dt\ga$ $3\times10^8$ g s$^{-1}$ display a warm infrared excess from 
orbiting dust; the likely result of a tidally-destroyed minor planet.  This benchmark accretion 
rate lies between the dust production rates of $10^6$ g s$^{-1}$ in the solar system zodiacal 
cloud and $10^{10}$ g s$^{-1}$ often inferred for debris disks at main sequence A-type stars.  
It is estimated that between 1\% and 3\% of all single white dwarfs with cooling ages less than 
around 0.5 Gyr possess circumstellar dust, signifying an underlying population of minor planets.

\end{abstract}


\section{INTRODUCTION}

Known extrasolar planetary systems orbiting main sequence stars consist of a few large 
planets such as Jupiter \citep{cum08}, and/or, as demonstrated by studies of debris disks, 
numerous minor planets analogous to solar system asteroids and Kuiper belt objects 
\citep{zuc01}.  Apparently, the assembly of planets from planetesimals is inefficient, and 
stars possess complex populations of orbiting material (see \citealt{ida08} and references
therein).

Relative to main sequence stars, white dwarfs offer two advantages for the study of extrasolar 
planetary systems. First, white dwarfs are earth-sized and their low luminosities permit the 
direct detection of infrared emission from cool self-luminous companions such as brown 
dwarfs and massive jovian planets \citep{far08a,bur06,far05a,zuc87a,pro83}.  Second, cool 
white dwarfs should be atmospherically free of heavy elements \citep{zuc03,alc86,paq86}, 
and those stars with planetary system remnants can become spectroscopically contaminated 
by small, but detectable, amounts of accreted material.  Analysis of metal-polluted white dwarfs 
enables an indirect, yet detailed and powerful compositional analysis of extrasolar planetary 
matter; \citet{zuc07} found that the abundances in the spectacularly metal-rich white dwarf 
GD 362 are consistent with the accretion of a large asteroid with composition similar to the 
Earth-Moon system.
 
The most metal-contaminated white dwarfs often display evidence of circumstellar disks; 
either by infrared excess \citep{far08b,kil07,von07,jur07a,kil06,bec05,kil05,gra90,zuc87b}, 
or by broad, double-peaked optical emission lines \citep{gan08,gan07,gan06}, or both 
\citep{mel08,bri08}.  Evidence is strong that these disks evolve from the tidal disruption 
of minor planets \citep{jur08,jur03}.  To be tidally destroyed within the Roche limit of a 
white dwarf, an asteroid needs to be perturbed from its orbit, and hence unseen planets 
of conventional size are expected at white dwarfs with dusty disks.

Including the disks around GD 16 and PG 1457$-$086, which are reported in this paper, 
the number of white dwarfs with circumstellar disks is 14 (see Table \ref{tbl1}).  Although 
at least half a dozen publications present {\em Spitzer} observations of white dwarfs, there 
has not yet been a thorough search of metal-rich white dwarfs for cool dust at longer, MIPS 
wavelengths, nor a focus on those contaminated degenerates with helium-rich atmospheres; 
this study bridges that gap and analyzes all 53 metal-rich degenerates observed by {\em 
Spitzer}.  The numbers are now large enough that one can investigate the presence of a 
disk as a function of cooling age and metal accretion rate.

This paper presents the results of an IRAC $3-8$ $\mu$m and MIPS 24 $\mu$m photometric 
search for mid-infrared excess due to circumstellar dust at cool, metal-contaminated white 
dwarfs.  The goals of the study are: 1) to constrain the frequency of dust disks around white 
dwarfs as a function of cooling age; and 2) to combine all available {\em Spitzer} data on 
metal-rich white dwarfs to better understand their heavy element pollutions.

\section{OBSERVATIONS AND DATA}

Metal-rich white dwarf targets were imaged over $3-8$ $\mu$m with the InfraRed Array Camera 
(IRAC, $1.20''$ pixel$^{-1}$; \citealt{faz04}) using all four bandpasses, and at 24 $\mu$m with 
the Multiband Imaging Photometer for {\em Spitzer} (MIPS, $2.45''$ pixel$^{-1}$ ; \citealt{rie04}). 
Observations analyzed here are primarily taken from {\em Spitzer} Cycle 3 Program 30387 and 
those previously published in \citet{jur07a}, with several archival datasets also included.  Twenty 
newly observed or analyzed white dwarf targets are listed in Table \ref{tbl2}, representing near 
equal numbers of DAZ and DBZ stars, and including G180-57 and HS 2253$+$803, two examples 
of metal-rich yet carbon-deficient stars.  Several Table \ref{tbl2} stars have been previously imaged 
with IRAC in other programs, hence for those stars only the MIPS observations are unique to this 
work; their IRAC data were extracted from the {\em Spitzer} archive and analyzed independently 
here.  For two metal-rich white dwarf targets of particular interest, vMa 2 and LTT 8452, both IRAC 
and MIPS data were extracted from the {\em Spitzer} archive.  The primary datasets utilized IRAC 
individual frame times of 30 s in a 20-point, medium-scale dither pattern, for a total exposure time 
of 600 s in each of the four filters at each star.  MIPS 24 $\mu$m observations were executed using 
10 s individual frame times with the default 14-point dither pattern, repeated for 10 cycles, yielding 
a 1400 s total exposure time for each science target.  Table \ref{tbl3} lists the IRAC and MIPS fluxes 
with errors and upper limits for all Table \ref{tbl2} white dwarfs, and also the non-metal-rich white 
dwarfs NLTT 3915 and LHS 46; these two stars are discussed in \S3.6.

Data reduction and photometry, including $3\sigma$ upper limits for non-detections, were 
performed as described in \citet{far08a,far08b}, and \citet{jur07a,jur07b}, but photometric errors 
were not treated as conservatively.  To account for the faintest MIPS 24 $\mu$m detections, and 
for reasons of consistency, the photometric measurement errors for both IRAC and MIPS were 
taken to be the Gaussian noise in an $r=2$ pixel aperture at all wavelengths; i.e. the sky noise 
per pixel multiplied by the square root of the aperture area.  IRAC data at 3.6 and 4.5 $\mu$m 
are typically high signal-to-noise (S/N $\ga50$) and therefore dominated by a 5\% calibration 
uncertainty \citep{far08b,jur07a}, whereas data at 5.7 and 7.9 $\mu$m have total errors which 
are a combination of calibration and photometric measurement errors.  Most of the MIPS 24 
$\mu$m detections are dominated by the photometric measurement errors, with a 10\% 
calibration uncertainty employed for these data \citep{eng07}.  In actuality, the photometric 
errors may be somewhat larger for reasons described in detail elsewhere; namely crowded 
fields for the two short wavelength IRAC channels, and non-uniform sky backgrounds for 
the two longer wavelength IRAC channels and for MIPS \citep{far08b}.  Among the 20 science 
targets imaged at 24 $\mu$m and analyzed here, the background noise varied significantly, 
as evidenced by the variance among the $1\sigma$ errors and $3\sigma$ upper limits in 
Table \ref{tbl3} (and similar variance among the 24 $\mu$m data in Table 1 of \citealt{jur07a}).

\section{ANALYSIS AND RESULTS}

Figures \ref{fig1}$-$\ref{fig8} plot the spectral energy distributions (SEDs) of the 20 metal-rich 
stars listed in Table \ref{tbl2}, grouped by right ascension, including {\em Spitzer} photometry 
and upper limits.  Optical and near-infrared photometric data were taken from: Table \ref{tbl2} 
references; \citet{mcc06} and references therein, \citet{sal03}, \citet{mon03}, and the 2MASS 
point source catalog \citep{skr06}.  The data were fitted with blackbodies of the appropriate 
temperature (i.e. able to reproduce the photospheric flux), which are sufficient to recognize 
excess emission at the $3\sigma$ level.  Of all the targets, only GD 16 and PG 1457$-$086 
display photometric excess above this threshold in their IRAC or MIPS observations.

\subsection{GD 16}

Figure \ref{fig2} displays the {\em Spitzer} photometry for GD 16, which reveals a prominent 
mid-infrared excess indicative of $T\sim1000$ K circumstellar dust.  The 2MASS data on this 
star appear to indicate an infrared excess at $H$ and $K$ bands, but it is likely these data are 
unreliable; independent near-infrared photometry at high S/N (Farihi et al. 2009, in preparation) 
indicate an excess only at $K$-band, as seen in Figure \ref{fig2}.  No published optical 
photoelectric photometry exists, but photographic data indicate $V\approx15.5$ mag from both 
USBO-B1 and the original discovery paper \citep{mon03,gic65}; this $V$-band flux is plotted 
and agrees well with the measured near-infrared fluxes for a white dwarf of the appropriate 
temperature.

A detailed optical spectral analysis of GD 16 by \citet{koe05b} yielded atmospheric parameters 
$T_{\rm eff}=11,500$ K, [H/He] $=-2.9$, and $M_V=12.05$ mag, with the assumption of log 
$g=8.0$.  Using the measured $J=15.55$ mag together with the model-predicted $V-J=0.09$ 
color for the relevant effective temperature, surface gravity, and helium-rich composition, the 
white dwarf is expected to have $M_J=12.14$ mag and hence a nominal photometric distance 
of $d=48$ pc \citep{ber95a,ber95b}.  Figure \ref{fig9} shows a fit to the thermal dust emission 
using the flat ring model of \citet{jur03}.  This model does an excellent job of reproducing all 
infrared data from 2.2 $\mu$m onwards.  For an 11,500 K star, the Figure \ref{fig9} inner and 
outer dust temperatures correspond to 12 and 30 stellar radii, or 0.15 and 0.39 $R_{\odot}$ 
respectively \citep{chi97}.  The fractional disk luminosity, $\tau=L_{\rm IR}/L=0.02$, is about 
$2/3$ that of the disks at G29-38, GD 362, and GD 56 \citep{far08b,jur07a}.  

\subsection{PG 1457$-$086}

{\em Spitzer} photometry for PG 1457$-$086 is shown in Figure \ref{fig5}, and reveals flux 
excess just over $3\sigma$ at both 3.6 and 4.5 $\mu$m, relative to the expected photospheric 
flux.  The 2MASS photometry for this white dwarf at $H$ and $K$ bands has large errors, and 
independent near-infrared photometry at high S/N (Farihi et al. 2009, in preparation) indicate 
a probable, slight excess at $K$ band.  Overall, the infrared excess is mild but indicative of 
very warm, $T\sim1500$ K emission.  \citet{koe05a} and \citet{lie05} derive $T_{\rm eff}=
20,400$ K and 21,500 K, respectively for PG 1457$-$086, while both find a surface gravity 
log $g\approx8.0$.  The 20,400 K value was employed for the figures because it is more 
conservative, predicting slightly less infrared excess than 21,500 K and because the model 
of \citet{koe05a} accounts for the measured high calcium abundance, while the \citet{lie05} 
model does not.

Figure \ref{fig10} shows a narrow ring model fitted to the infrared excess.  To match the 
relatively low fractional luminosity of the disk, $\tau=0.0006$, the model ring is narrow in 
radial extent and highly inclined. For a 20,400 K white dwarf, the inner and outer dust 
temperatures, in an optically thick disk, correspond to 19 and 21 stellar radii, or 0.25 
and 0.28 $R_{\odot}$ respectively \citep{chi97}.

Due to the shape of the mild infrared excess at PG 1457$-$086, it is the only metal-rich 
white dwarf {\em Spitzer} target where a substellar companion might be considered as 
viable.  If the measured $J=16.07$ mag is solely due to the white dwarf photosphere, 
then the difference between the measured and model-predicted $K$-band flux for such 
a 20,000 K hydrogen white dwarf is $\Delta K=0.24$ mag or an excess of $K=17.8$ mag, 
which corresponds to $M_K=12.6$ mag at the photometric distance of 110 pc \citep{lie05}.  
If a self-luminous, substellar companion of this $K$-band brightness were orbiting PG 
1457$-$086, it would have an effective temperature near 1500 K and a spectral type 
around L7 \citep{vrb04,dah02}.  It is worth noting that the white dwarf GD 1400 has an 
unresolved L7 brown dwarf companion, and the ratio of the 3.6 to 2.2 $\mu$m excess 
there is 1.3 \citep{far05b}, whereas for PG 1457$-$086 this value is 0.7, although these 
ratios are consistent within the uncertainties.  

As mentioned in the Appendix, winds from nearby M dwarf companions can, and occasionally 
do, pollute the atmospheres of white dwarfs.  But for PG 1457$-$086 an M dwarf companion is 
ruled out by the absence of significant near-infrared excess \citep{far05a}.  As described in the 
Appendix, there is no known connection between brown dwarf companions and the presence 
of metal pollution in white dwarf photospheres.  Therefore, for PG 1457$-$086 to be metal-rich 
and to possess an L dwarf companion would require, simultaneously, two low probability events.  
For this reason, it is it highly unlikely that the excess infrared emission at PG 1457$-$086 is due 
to a brown dwarf companion.

\subsection{LTT 8452}

\citet{von07} reported the discovery of {\em Spitzer} IRAC 4.5 and 7.9 $\mu$m excess due 
to warm dust at LTT 8452.  Figure \ref{fig7} shows the SED of this metal-rich white dwarf, 
now plotted with its IRAC 3.6, 5.7, and MIPS 24 $\mu$m data, which better constrain the 
inner and outer temperature of the disk.  Figure \ref{fig11} shows a flat ring model fitted to 
the mid-infrared emission of LTT 8452, with inner edge temperature $T_{\rm in} =1000$ K, 
outer temperature $T_{\rm out}=600$ K, and a modest inclination angle of $i=53 \arcdeg$,
whereas \citet{von07} derived a temperature range of $900-550$ K and $i=80 \arcdeg$.  
From the flat disk model, the fractional infrared luminosity of LTT 8452 is $\tau= 0.008$.

\subsection{G238-44 and G180-57}

Figure \ref{fig4} includes all available {\em Spitzer} data for G238-44, one of the most 
highly contaminated, nearest, and brightest DAZ white dwarfs \citep{hol97}.  The aperture 
laid down for MIPS 24 $\mu$m photometry was extrapolated to the expected location of the 
star at the epoch 2007.3 observation, using the {\em Hipparcos} measured J2000 position 
and proper motion \citep{per97}.  The expected position on the MIPS 24 $\mu$m array is in 
excellent agreement with the measured position of G238-44 in the epoch 2004.9 IRAC 
images, after accounting for proper motion, and coincides with a faint MIPS 24 $\mu$m 
source.  However, the potential for source confusion is high.

If associated with the white dwarf, the Figure \ref{fig4} apparent 24 $\mu$m excess at the 
location of G238-44 would be just slightly greater than 0.04 mJy ($2.0\sigma$).  Based on 
{\em Spitzer} MIPS 24 $\mu$m source counts from deep imaging, the expected number of 
background galaxies of brightness $0.04\pm0.02$ mJy is around 8000 per square degree 
\citep{mar04}.  Hence within an area of diameter equal to one full width at half maximum 
intensity at this wavelength (about 2.3 pixels or approximately 25 square arcseconds), there 
should be 0.015 galaxies of the right brightness to contaminate the MIPS aperture.  Using a
binomial probability, the chance of finding at least one in 20 metal-rich white dwarf targets 
confused with a background source in this manner is then 26\%; therefore the 24 $\mu$m 
flux may not originate from G238-44.

Further, if the potential contamination area is widened to encompass two full widths at half 
maximum, the probability of source confusion is then 71\%, although such a large an area 
is perhaps overly conservative.  The expected position of G238-44 is offset from the centroid 
position of the MIPS source by 0.35 pixels or $0.9''$, the error of which is hard to estimate 
due to the relatively low S/N.

Turning to the white dwarf G180-57 (Figure \ref{fig4}), it has a 0.06 mJy (or $1.5\sigma$ excess) 
24 $\mu$m source at its expected position on the array.  All the previous arguments and caveats 
apply, and the chance that at least two in 20 MIPS targets are contaminated in this way is still 
relatively high, somewhere between 4\% and 34\%.

In any event, cold dust alone at this 24 $\mu$m luminosity level cannot explain the source 
of external metals in these white dwarfs.  Blackbody dust which reveals itself at 24 $\mu$m but 
not at 8 $\mu$m would be around 200 K or cooler, and located too far away (at or beyond 10 
and 70 $R_{\odot}$ for G180-57 and G238-44 respectively) to be the source of accreted metals 
in the photospheres of either white dwarf.  Average-sized grains of 10 $\mu$m with density 2.5 g 
${\rm cm}^{-3}$ at these distances would imply (minimum) dust masses around $10^{15}-10
^{17}$ g for $\tau=10^{-5}$, appropriate for these detections, if real \citep{far08b}.  This mass is 
insufficient to sustain an accretion rate of $3\times10^8$ g ${\rm s}^{-1}$ for more than 10 years 
at G238-44 \citep{koe06}.

\subsection{vMa 2}

\begin{figure}
\plotone{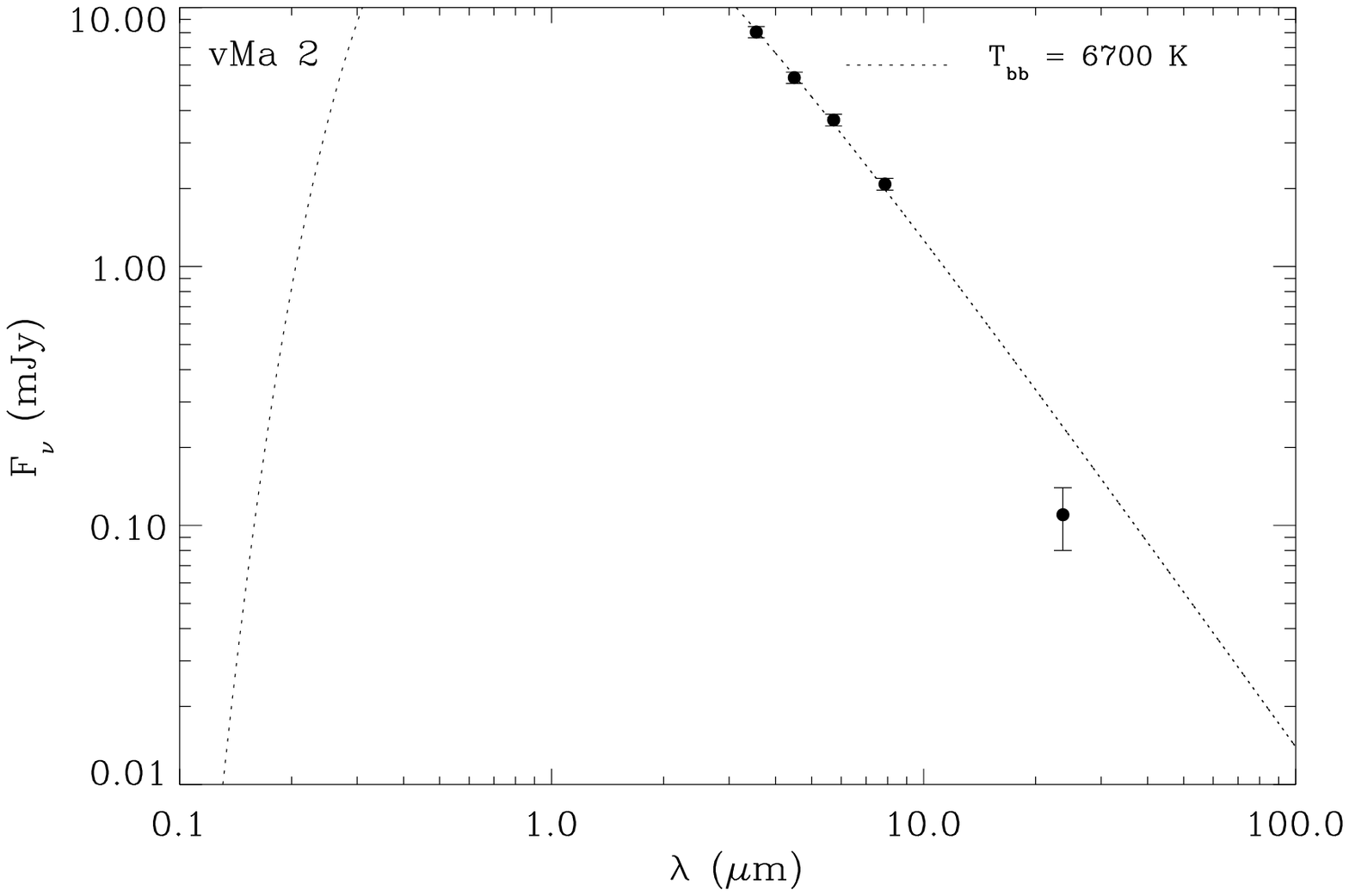}
\caption{SED of vMa 2.
\label{fig1}}
\end{figure}

\begin{figure}
\plotone{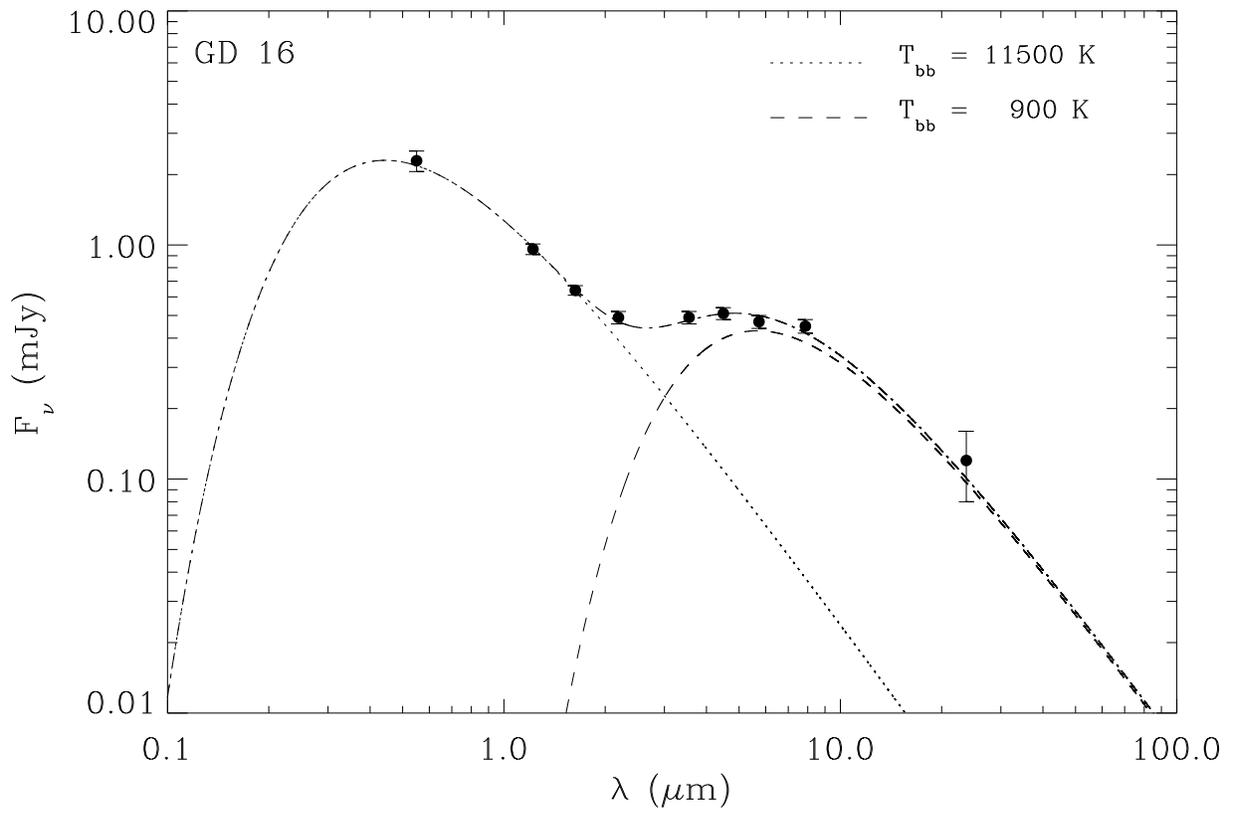}
\caption{SED of GD 16.
\label{fig2}}
\end{figure}

\begin{figure}
\plotone{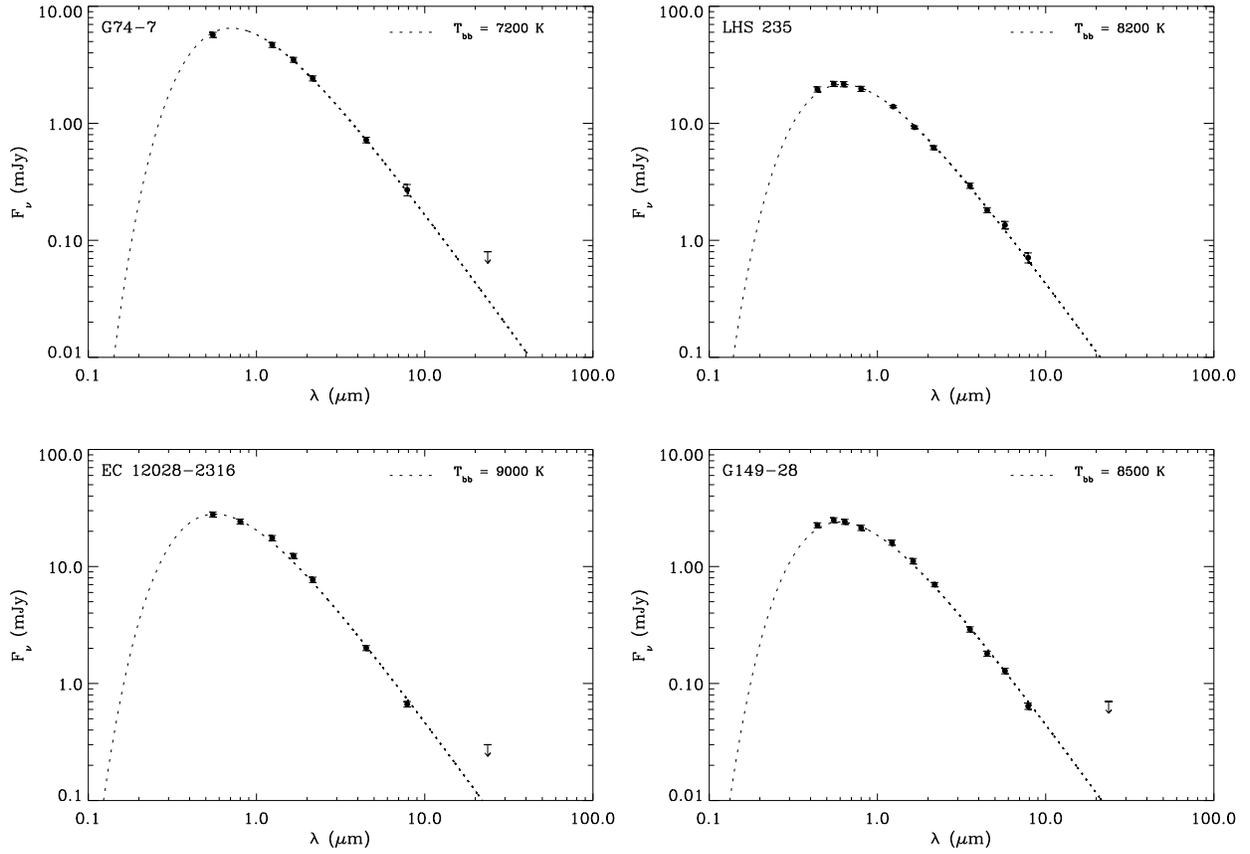}
\caption{SEDs of G74-7, LHS 235, EC 1202$-$232, and G149-28.  Downward arrows 
represent $3\sigma$ upper limits (\S2).
\label{fig3}}
\end{figure}

\begin{figure}
\plotone{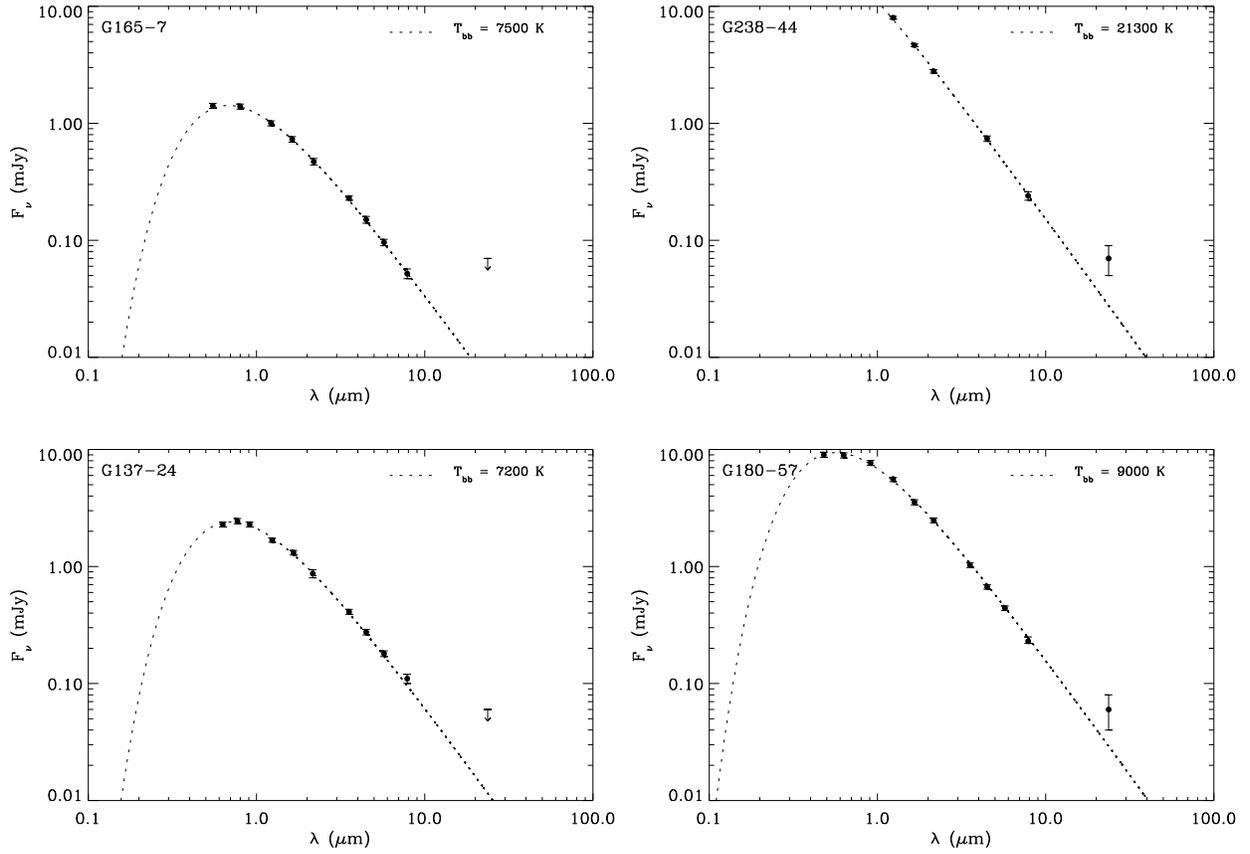}
\caption{SEDs of G165-7, G238-44, G137-24, and G180-57.  Downward arrows represent 
$3\sigma$ upper limits (\S2).
\label{fig4}}
\end{figure}

\begin{figure}
\plotone{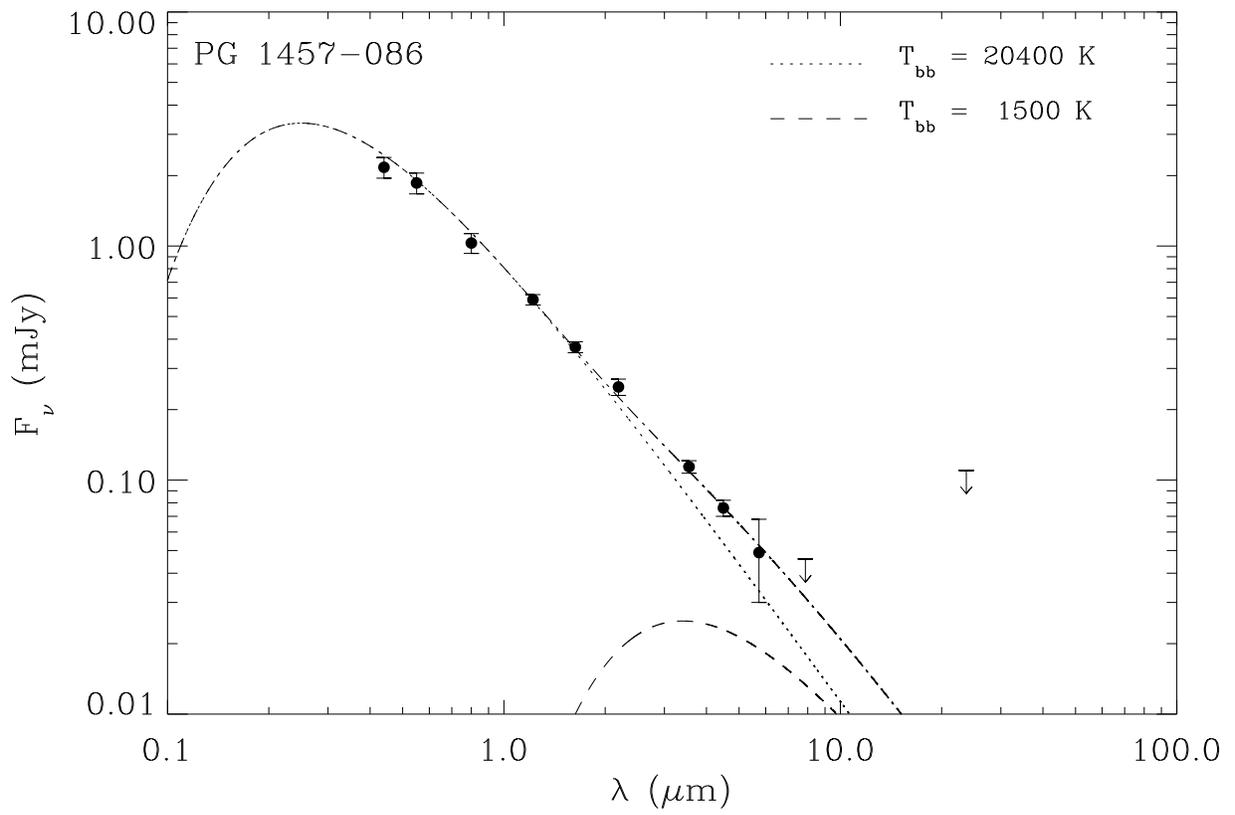}
\caption{SED of PG 1457$-$086.  Downward arrows represent $3\sigma$ upper limits (\S2).
\label{fig5}}
\end{figure}

\begin{figure}
\plotone{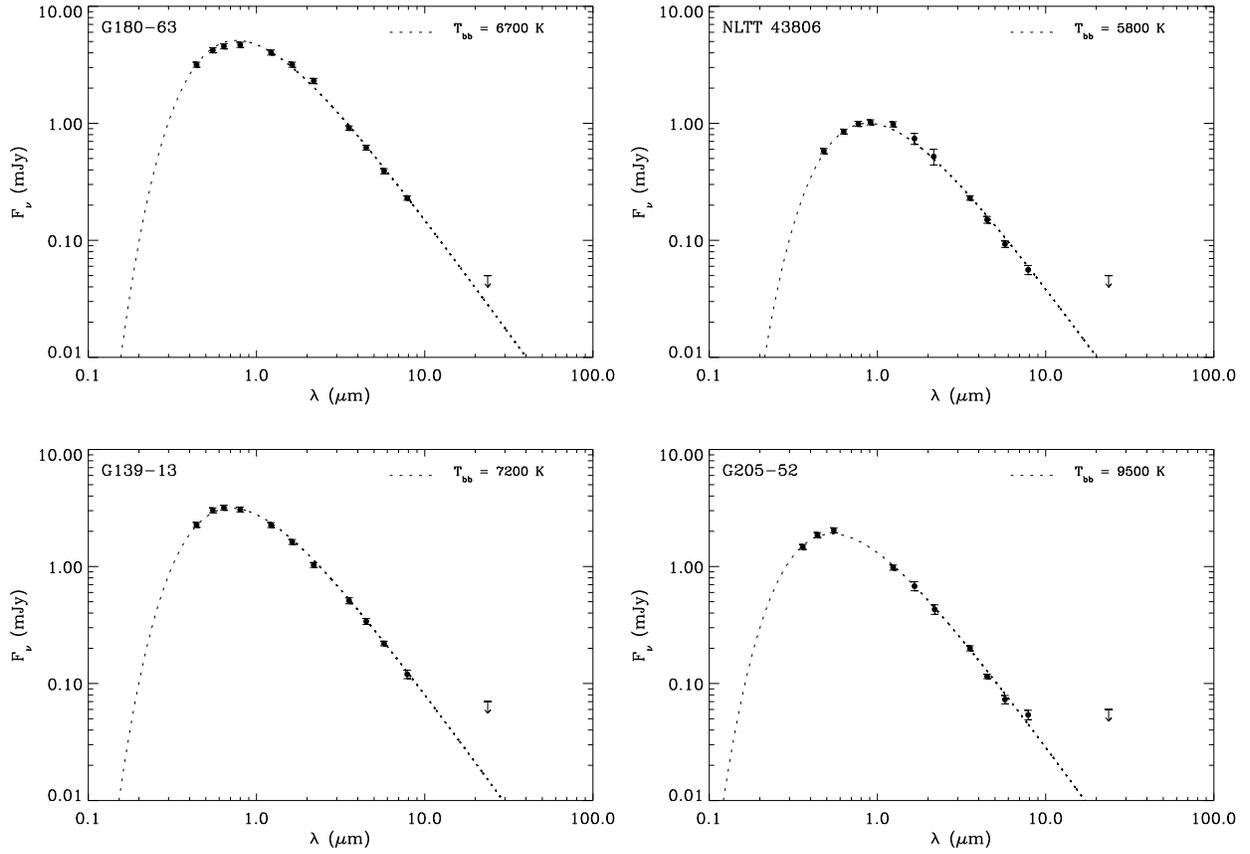}
\caption{SEDs of G180-63, NLTT 43806, G139-13, and G205-52.  Downward arrows 
represent $3\sigma$ upper limits (\S2).
\label{fig6}}
\end{figure}

\begin{figure}
\plotone{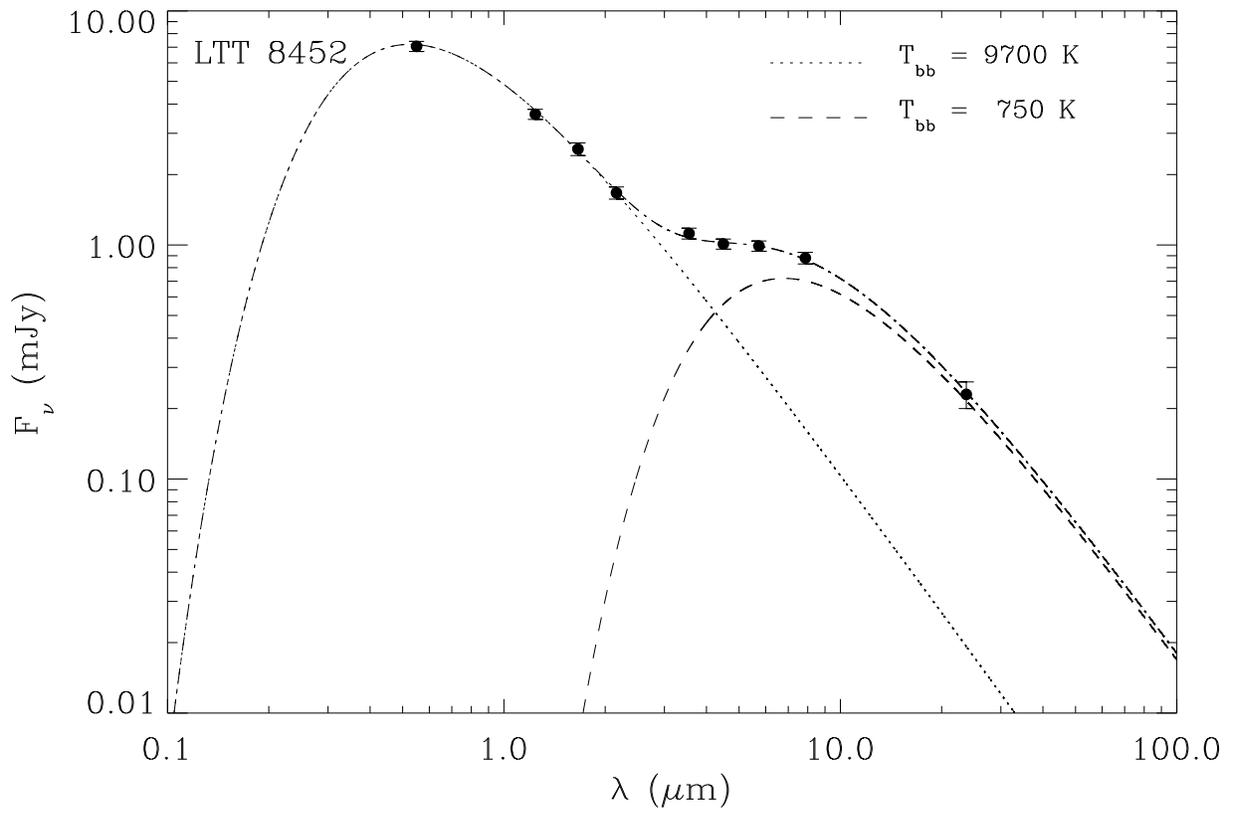}
\caption{SED of LTT 8452.
\label{fig7}}
\end{figure}

\begin{figure}
\plotone{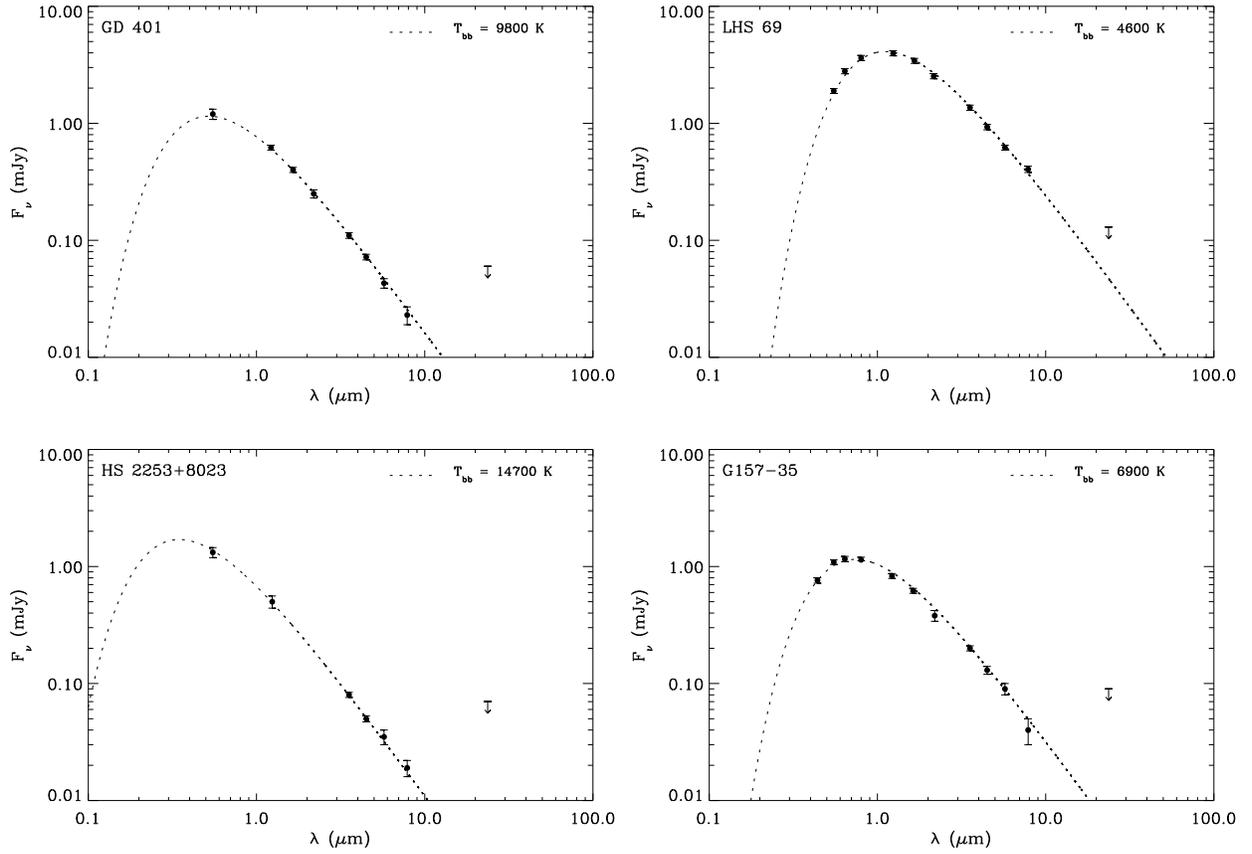}
\caption{SEDs of GD 401, LHS 69, HS 2253$+$8023, and G157-35.  Downward arrows 
represent $3\sigma$ upper limits (\S2).
\label{fig8}}
\end{figure}

\begin{figure}
\plotone{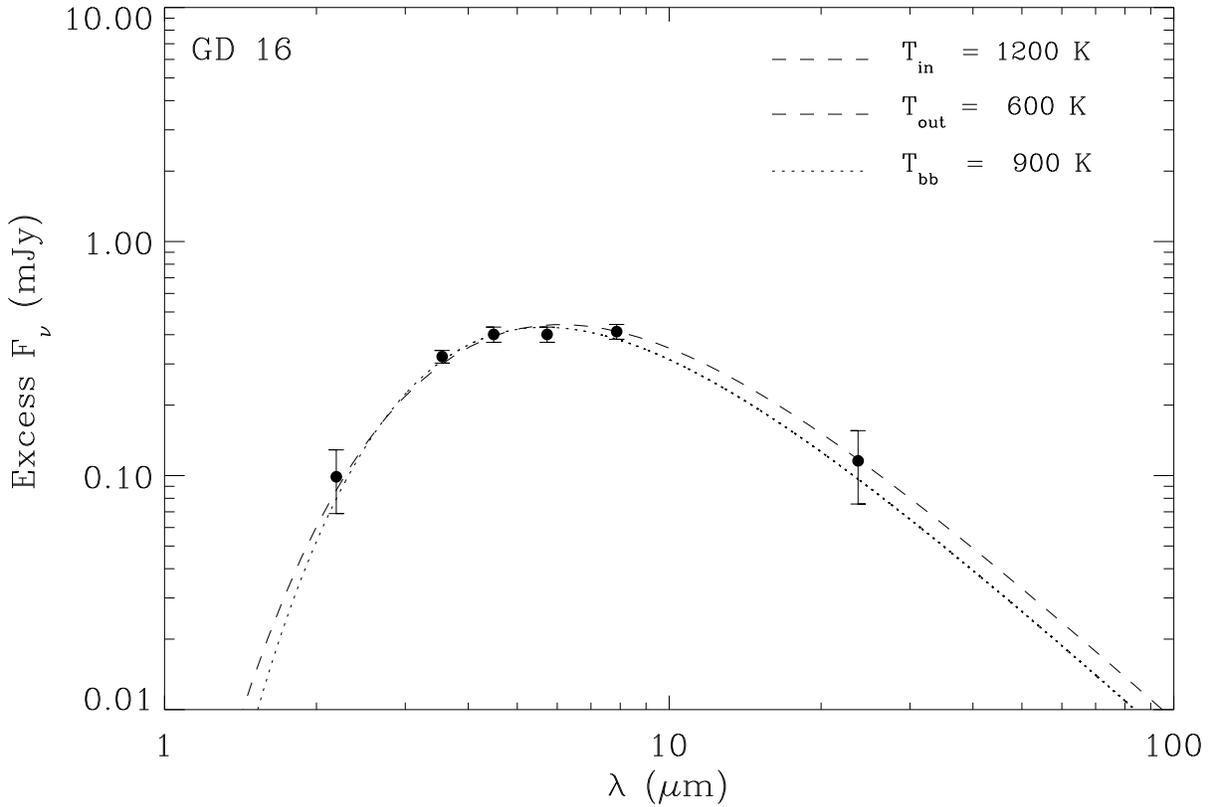}
\caption{Excess infrared emission above the stellar photosphere from GD 16 fitted with a 
model circumstellar disk.  The dashed line represents an optically thick, geometrically thin 
disk of inclination angle $i=48\arcdeg$ ($i=0\arcdeg$ is face-on) and finite radial extent, 
whose inner edge temperature is $T_{\rm in}=1200$ K and whose outer temperature is 
$T_{\rm out}=600$ K \citep{jur03}.  As may be seen, emission from a 900 K blackbody is 
a close approximation to the model ring emission.
\label{fig9}}
\end{figure}

\begin{figure}
\plotone{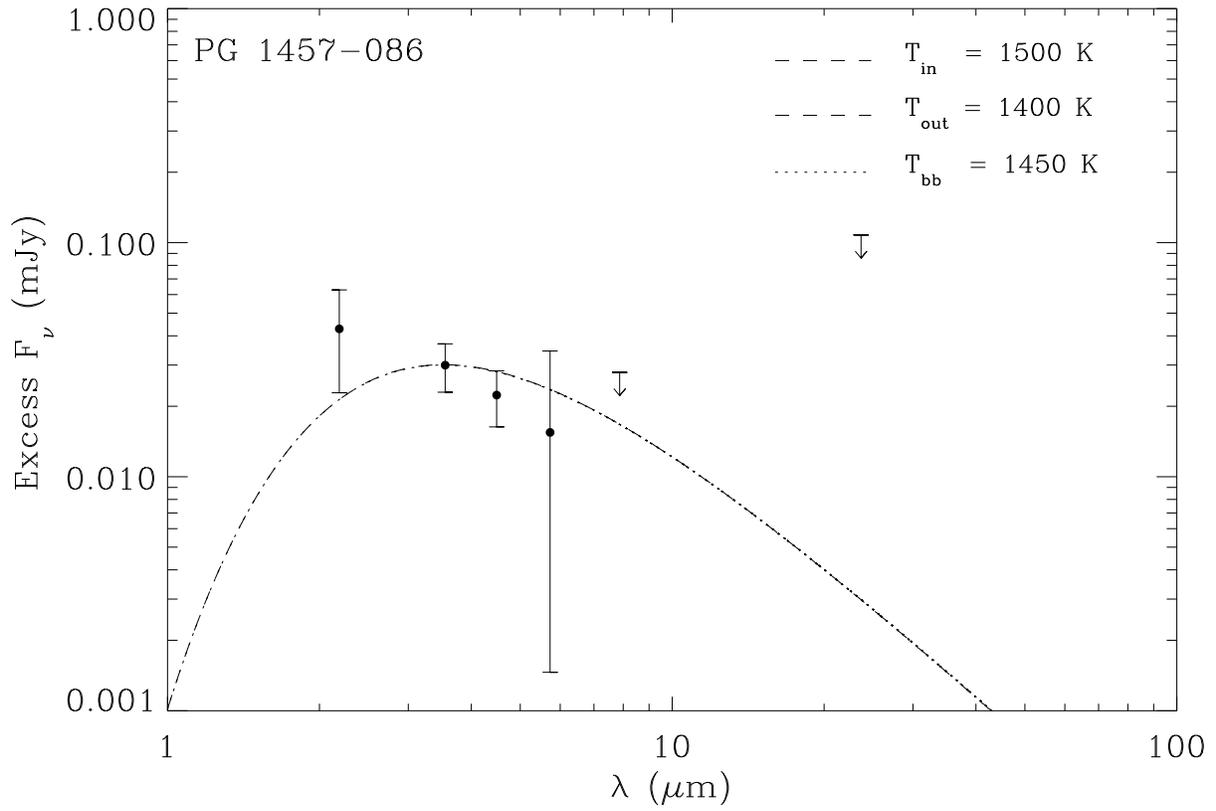}
\caption{Excess infrared emission above the stellar photosphere from PG 1457$-$086 fitted 
with a model circumstellar disk with $i=73\arcdeg$ (see Figure \ref{fig9} caption).  The 1450 
K blackbody well approximates the optically thick, narrow ring.
\label{fig10}}
\end{figure}

\begin{figure}
\plotone{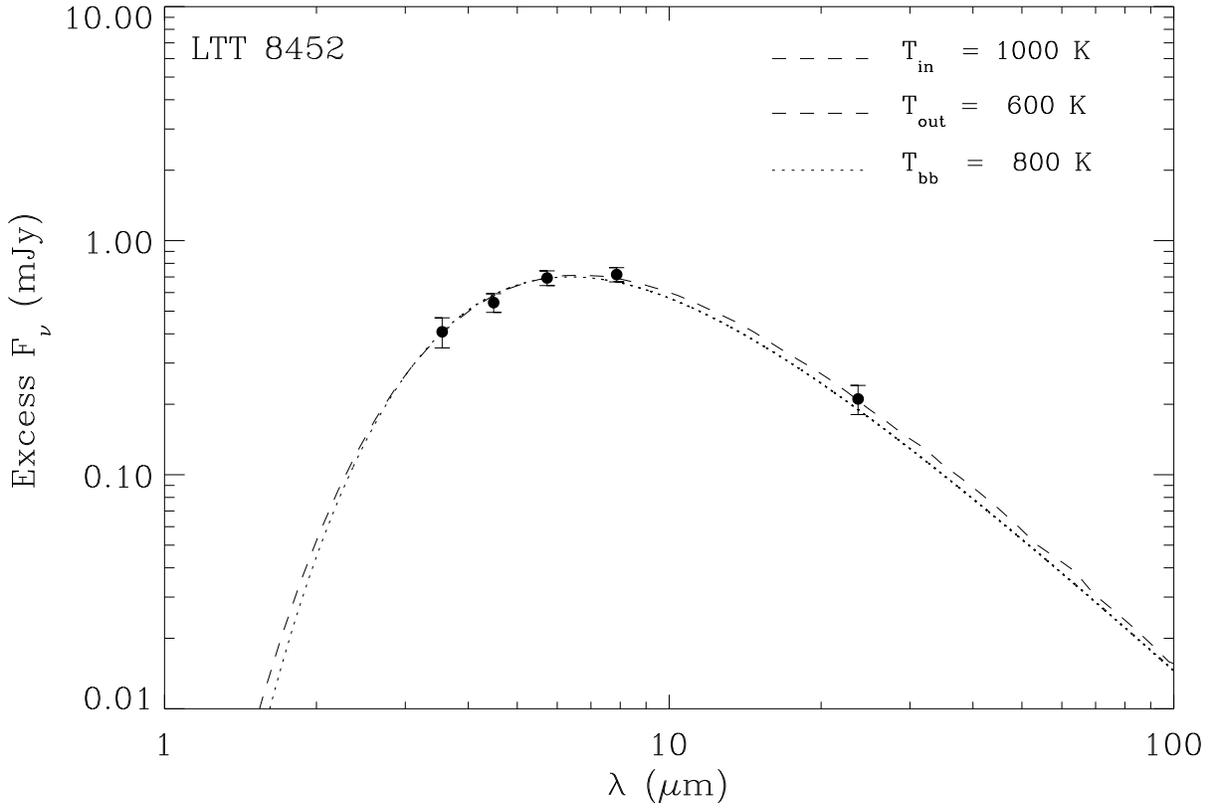}
\caption{Excess infrared emission above the stellar photosphere from LTT 8452 fitted with a 
model circumstellar disk.  The 800 K blackbody approximates the optically thick, narrow ring 
with a temperature range of $1000-600$ K.
\label{fig11}}
\end{figure}

The MIPS observations of vMa 2, plotted in Figure \ref{fig1}, suggest a 24 $\mu$m flux 
significantly lower than expected for a simple Rayleigh-Jeans extrapolation from its IRAC 
fluxes.  Rather than an excess, the SED of vMa 2 appears deficient at 24 $\mu$m, at the 
$4\sigma$ level; the measured 24 $\mu$m flux is $0.11\pm0.03$ mJy, while the predicted 
flux is 0.23 mJy (Figure \ref{fig1}).  Blackbody models are essentially no different than pure 
helium atmosphere white dwarf models at these long wavelengths (\citealt{wol02}; see their 
Figure 1), hence the apparent deficit rests on the validity of the relatively low S/N photometry 
rather than model predictions.  Still, this intriguing possibility requires confirmation with 
superior data, and if confirmed, vMa 2 would become by far the highest temperature white 
dwarf to display significant infrared flux suppression due to collision-induced absorption 
(B. Hansen 2007, private communication; \citealt{far05c}).

\subsection{Notes on Individual Objects}

{\em 0108$+$277}.  NLTT 3915 belongs to an optical pair of stars separated by roughly $3''$ 
on the sky in a 1995.7 POSS II plate scan.  The northeast star has proper motion $\mu=0.22''$ 
yr$^{-1}$ at $\theta=222\arcdeg$ \citep{lep05} which can be readily seen between the POSS I 
and POSS II epochs.  The IRAC images reveal two overlapping sources with a separation of 
$2.4''$ as determined by {\sf daophot} at all four wavelengths.  Using this task to deconvolve 
the pair of stars photometrically reveals the object to the southwest is likely a background red 
dwarf, based on its 2MASS and IRAC photometry, while the northeast star has colors consistent 
with a very cool white dwarf.  \citet{kaw06} identified this star as a 5200 K DAZ white dwarf 
whose spectrum appears to exhibit sodium but not calcium; an anomalous combination for 
a metal-rich degenerate.  Keck / HIRES observations confirm this object is a white dwarf, but 
with a cool DA spectrum and no metal features (C. Melis 2008, private communication).  The 
data tables include fluxes, and the appendix gives limits on substellar companions for this 
target, but it is excluded it from analyses for metal-rich white dwarfs.

{\em 0208$+$396}. G74-7 is the prototype DAZ star.  IRAC observations of this white dwarf, 
previously reported in \citet{deb07} were taken during a solar proton event, and individual 
frames are plagued by legion cosmic rays, making the reduction and photometric analysis 
difficult, especially at the longer wavelengths.  Only the 4.5 and 7.9 $\mu$m fluxes are plotted 
and tabled; some data were irretrievably problematic.

{\em 1202$-$232}.  This bright and nearby white dwarf is located about $8''$ away from 
a luminous infrared galaxy in the IRAC 8 $\mu$m images, and the white dwarf location is 
swamped by light from the galaxy at MIPS 24 $\mu$m, where the diffraction limit is $6''$.  
Hence the upper limit on the white dwarf 24 $\mu$m flux is about a factor of five worse than 
for a typical star.

{\em 1334$+$039}.  LHS 46 has been (mistakenly) identified as spectral type DZ in several 
papers over the years (\citealt{gre84}; see discussion in \citealt{lie77}).  It was first (correctly) 
re-classified as a DC star by \citet{sio90}, and modern, high resolution spectroscopy confirms 
the white dwarf is featureless in the region of interest \citep{zuc98}.  Unfortunately, some 
relatively current literature \citep{hol08,hol02,mcc99} still contains the incorrect spectral 
type for this star and it was included in the {\em Spitzer} observations.  The data tables 
include fluxes, and the appendix gives limits on substellar companions for this target, 
but it is excluded it from analyses for metal-rich white dwarfs.

\section{DISCUSSION}

With the detection of mid-infrared excess at GD 16 and PG 1457$-$086, the number of 
externally polluted, cool white dwarfs with warm circumstellar dust becomes 14, including 
EC 1150$-$153 \citep{jur09}, SDSS 1228 \citep{bri08} (see Table \ref{tbl1}), SDSS 1043 
and Ton 345 (C. Brinkworth 2008, private communication; \citealt{mel08}).  The latter three 
stars, whose photospheric metals and circumstellar gas disks were discovered simultaneously, 
are not analyzed here.  Over 200 white dwarfs have been observed with {\em Spitzer} IRAC 
\citep{far08a,far08b,mul07,deb07,han06,jur07b,fri07}.  Of these, only stars with detected 
photospheric metals display an infrared excess, presumably because metal pollution from 
a circumstellar disk is inevitable and optical spectroscopy is a powerful tool for detecting 
the unusual presence of calcium in a white dwarf atmosphere.  While the sample of white
dwarfs observed with {\em Spitzer} is quite heterogeneous, it is likely that at least half of 
these stars have been observed in the Supernova Progenitor Survey (SPY; \citealt{nap03}),
and hence with high sensitivity to photospheric calcium.

\subsection{The Fraction of Single White Dwarfs with an Infrared Excess}

The fraction of white dwarfs cooler than about 20,000 K that display photospheric metals 
depends on the stellar effective temperature and on whether its photospheric opacity is 
dominated by hydrogen or by helium.  For DA white dwarfs, it is easier to detect metals in 
cooler stars and, in a survey that focused primarily on DA white dwarfs cooler than 10,000 
K, \citet{zuc03} found that of order 25\% displayed a calcium K line.  In the extensive SPY 
survey, more focused on warmer white dwarfs with $T_{\rm eff}\ga10,000$ K, \citet{koe05a} 
found that only 5\% show a calcium K line.  To produce a detectable optical wavelength line 
at such high temperatures (and high opacities) typically requires a greater fractional metal 
abundance than at low temperatures.  Because extensive ground and, especially, {\em 
Spitzer} surveys have failed to reveal infrared excess emission at any single white dwarf 
that lacks photospheric metals \citep{far08b,mul07,hoa07,han06,far05a}, one can regard 
the above percentages as upper limits to the fraction of DA white dwarfs that possess dusty 
disks; at least until more sensitive metal abundance measurements can be achieved.  
Perhaps more instructive is the number DA stars without metals observed with {\em 
Spitzer}; 121 such targets are reported between \citet{mul07} and \citet{far08a}, none 
of which show evidence for circumstellar dust, an upper limit of 0.8\%.

Lower limits to the fraction of DA white dwarfs with dust disks can be estimated in the following 
way.  Consider four large surveys of white dwarfs: the Palomar-Green Survey (347 DA stars;
\citealt{lie05}), the Supernova Progenitor Survey (478 DA stars; \citealt{koe05a}); the 371
white dwarfs from \citet{far05a}, and the 1321 non-binary stars present in both the 2MASS
\citep{hoa07} and \citet{mcc99} catalogs.  The presently known frequency of dust disks 
for white dwarfs in these four surveys, all of which tend to find warmer white dwarfs (due
to luminosity), are 1.4, 1.5, 1.1 and 0.8\%, respectively.  Thus, because not all stars in these 
surveys have been searched for dust disks, one can say that, at a minimum, 1\% of all warm 
white dwarfs have dust disks. 

{\em Spitzer} surveys of cool metal-rich white dwarfs have now targeted 53 stars at $3-8$ 
$\mu$m using IRAC (a few at 4.5 and 7.9 $\mu$m only, but most at all wavelengths), while 
31 of these targets have also been observed at 24 $\mu$m with MIPS (see Table \ref{tbl4}).  
Including the unlikely exceptions discussed above, there does not exist a clear-cut case of 
MIPS 24 $\mu$m detection at any white dwarf without a simultaneous and higher IRAC flux 
-- a telling result by itself.  Hence, targets observed with IRAC only should accurately reflect 
the frequency of dusty degenerates.  

\begin{figure}
\plotone{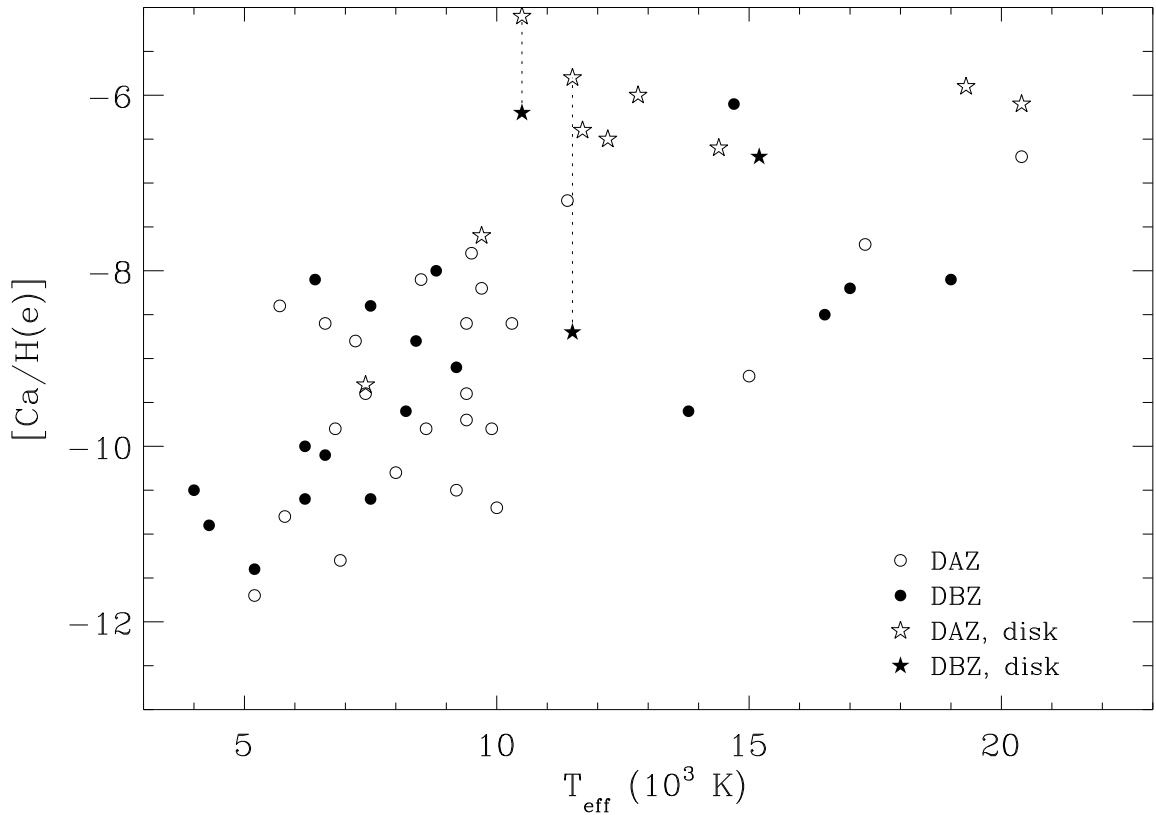}
\caption{Calcium to hydrogen (for DAZ stars) and calcium to helium (for DBZ stars) abundance 
ratios for all metal-contaminated white dwarfs studied by {\em Spitzer} IRAC.  In the plot, GD 16 
and GD 362, both of spectral type DAZB (Table \ref{tbl4}), are plotted both as DAZ and as DBZ 
stars, their data points connected by dotted lines.  A typical uncertainty in the calcium abundance
is 0.1 dex.
\label{fig12}}
\end{figure}

Of the 53 contaminated white dwarfs surveyed specifically for circumstellar disks, 21\% 
have mid-infrared data consistent with warm dust, regardless of atmospheric composition.
With more data, this estimate is somewhat larger than the result of \citet{kil08} that at least 
14\% of polluted white dwarfs have an infrared excess.  As shown in Figure \ref{fig12}, the 
likelihood of a white dwarf displaying an infrared excess is strongly correlated with effective 
temperature and/or its measured calcium pollution.  Only two of 34 stars with $T_{\rm eff}\leq
10,000$ K have an excess, while that fraction is nine of 19 stars with $T_{\rm eff}>10,000$ 
K.  Alternatively, nine (or ten) of 17 stars with [Ca/H(e)] $\geq-8.0$ possess an excess, 
whereas that fraction is only one (or two) of 38 stars with [Ca/H(e)]  $<-8.0$.

\subsubsection{The Role of Cooling Age}

Cooling ages for the white dwarfs in Table \ref{tbl4} were calculated with the models of P. 
Bergeron (2002, private communication; \citealt{ber95a,ber95b}). The white dwarf masses 
used as input for the cooling ages were taken from the literature (see Table \ref{tbl4} references), 
or log $g=8.0$ was assumed for those stars with no estimates available.  Because cooling age is 
sensitive to mass, and because within the literature discrepancies exist among white dwarf mass 
estimates, the values here should not be considered authoritative, but preference was given to 
values based on trigonometric parallaxes \citep{hol08} and those with multiple, independent,
corroborating determinations.  Still, it should be the case that a typical uncertainty in cooling 
age is between 10\% and 20\% due to an error in white dwarf mass.

\begin{figure}
\plotone{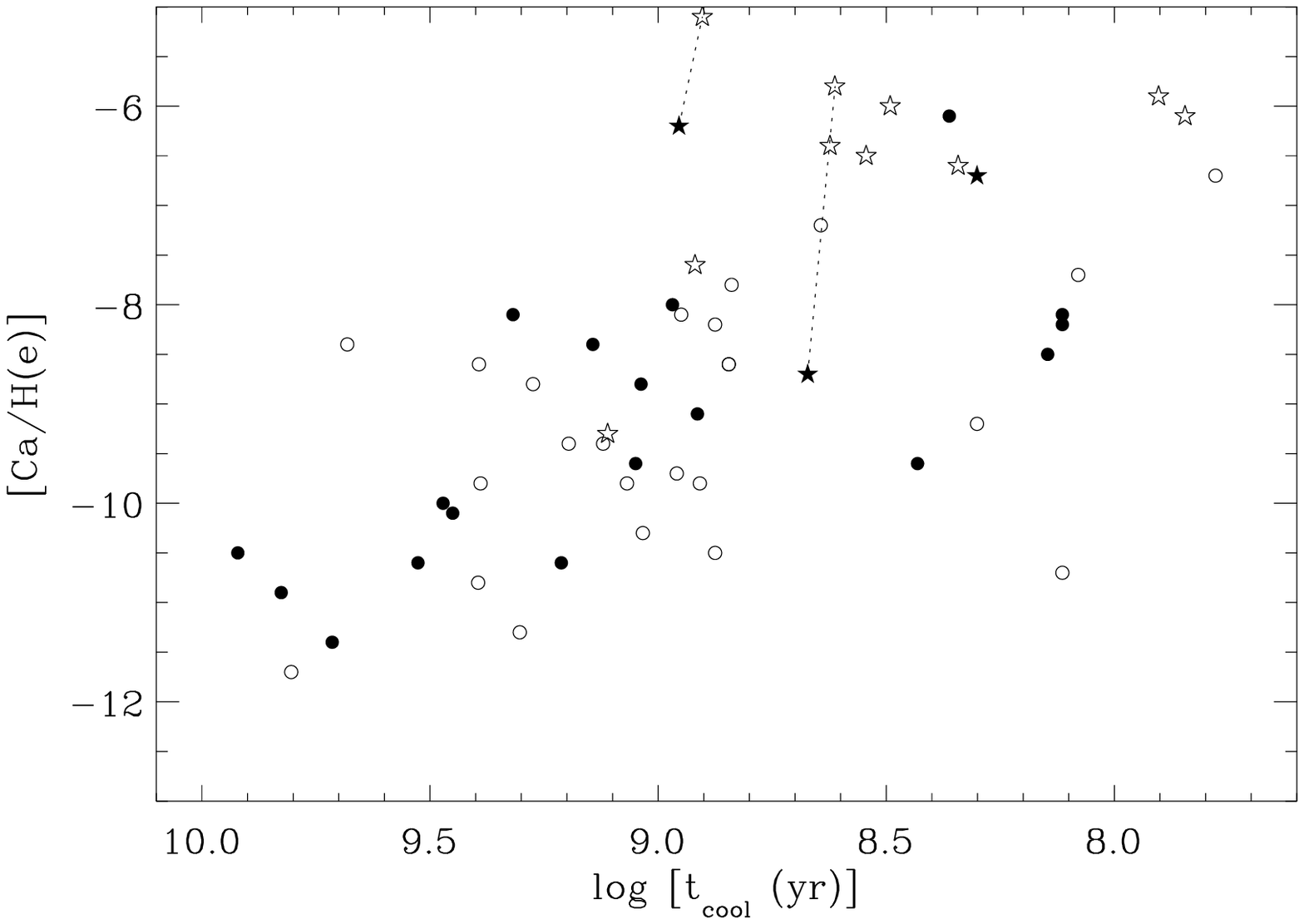}
\caption{Same as Figure \ref{fig12} but plotted versus cooling age.
\label{fig13}}
\end{figure}

Figure \ref{fig12} indicates that the most metal-polluted white dwarfs are the warmest, and 
Figure \ref{fig13} illustrates the same phenomenon, but now cast in terms of white dwarf 
cooling age.  In the context of a model of tidal destruction of minor planets, as described 
below, it is plausible that younger (i.e., shorter cooling time) white dwarfs would be the most 
polluted.  As indicated in the figure and tables, circumstellar dust disks are found at:  eight of 
17 stars (47\%) with $t_{\rm cool}<0.5$ Gyr, two of 12 stars (17\%) with 0.5 Gyr $<t_{\rm cool}
<1.0$ Gyr, and only one of 24 (4\%) stars with $t_{\rm cool}>1.0$ Gyr.  The three dusty white 
dwarfs with cooling ages beyond 0.5 Gyr are: GD 362 and LTT 8452, both with cooling ages 
near 0.8 Gyr; and G166-58 with cooling age 1.29 Gyr.

To estimate the percentage of white dwarfs with cooling ages less than 0.5 Gyr that might have 
dusty disks, one can consider each of the four large surveys mentioned in the previous section.
For example, the SPY survey contains 14 metal-contaminated DA white dwarfs with likely cooling 
ages less than 0.5 Gyr ($T_{\rm eff}\ga11,000$ K).  Of these 14, eight have been observed with 
IRAC and six have dust disks, while the remaining six, unobserved stars are prime candidates 
for dust disks (see \S4.2).  The SPY sample contains approximately 400 DA stars in this range
of cooling ages and therefore, based on these data, it is likely that between 2\% and 3\% of white 
dwarfs with $t_{\rm cool}<0.5$ Gyr have detectable dust disks.  This estimate is also consistent 
with the 72 DB white dwarfs (all with $T_{\rm eff}\ga11,000$ K) observed with SPY \citep{vos07,
koe05b}, two of which are metal-contaminated and possess dust disks.

In a similar manner, the SPY survey contains 10 metal-contaminated DA white dwarfs with 
likely cooling ages between 0.5 and 1.5 Gyr (7000 K $\la$ $T_{\rm eff}\la11,000$ K).  Of these
10, eight have been observed with IRAC and one has a dust disk, while the remaining two,
unobserved stars both have [Ca/H] $>-8.0$, and hence one of them may have a disk.  The 
SPY sample contains around 70 DA stars in this range of cooling ages, and therefore these 
data suggest between 1\% and 2\% white dwarfs with 0.5 Gyr $<$ $t_{\rm cool}<1.5$ Gyr 
may have detectable dust disks.  However, the smaller number statistics for these cooler
white dwarfs make this estimate somewhat uncertain.

Assuming only metal-bearing white dwarfs can possess warm dust disks, the {\em Spitzer}
observations suggest similar percentages.  The fraction of younger, $t_{\rm cool}<0.5$ Gyr 
white dwarfs with dust disks can be estimated by observing the fraction of these with IRAC 
excess is 0.47, and the fraction of SPY DA white dwarfs with metals in this cooling age range 
is 0.05, leading to a frequency between 2\% and 3\%, consistent with the above estimate.  For 
somewhat older white dwarfs where 0.5 Gyr $<t_{\rm cool}<1.5$ Gyr, the same fractions are 
0.10 with IRAC excess, and 0.13 with metal lines among SPY DA targets, yielding a frequency 
of 1\%.  For white dwarfs with $t_{\rm cool}>1.5$ Gyr, there there are not enough stars in 
appropriate age bins to make meaningful estimates.

\citet{hol08} have compiled a nearly (estimated at 80\%) complete catalog of white dwarfs 
within 20 pc of the Sun.  In this local volume, there are 24 white dwarfs with 10,000 K $<
T_ {\rm eff}\la 20,000$ K, and at least one white dwarf (G29-38), i.e. 4\%, has an infrared 
excess.  If the true percentage of warm dusty white dwarfs is 2.5\%, then out to 50 pc there 
should be nine such stars with infrared excess.  Currently, only GD 16 and GD 133 meet 
these criteria; either this expectation is incorrect or several white dwarfs within 50 pc of 
the Sun with an infrared excess remain to be discovered.

\subsubsection{Implications for Disk Lifetimes and Pollution Events}

The relative dearth of dust disks at the cooler metal-rich white dwarfs, as established by 
{\em Spitzer} observations, suggests that dust disk lifetimes are likely to be shorter than 
a typical $10^4-10^6$ yr heavy element diffusion timescale in either a hydrogen or helium 
atmosphere white dwarf at these temperatures.  Whether a typical disk mass is fully consumed 
within that period or is rendered gaseous via collisions on much shorter timescales is unclear, 
and both mechanisms are likely to play a role to remove dust at white dwarfs \citep{jur08,
far08b,jur07a}.  However, as noted above, the lack of dust at the cooler polluted stars is also 
related to their lower overall metal abundances, and partially an observational bias.

More important, this empirical result may have implications for the timescales of the 
events which give rise to and/or sustain circumstellar dust.  If stochastic perturbations within 
a reservoir of planetesimals are the ultimate source of pollution events in metal-enriched 
white dwarfs, gradual depletion could give rise to a exponential decay in the number of 
events per unit time \citep{jur08}, and hence the likelihood of a pollution event decreases 
as a white dwarf ages.  Such a scenario is consistent with the {\em Spitzer} observations 
of the cooler, metal-lined white dwarfs, and would predict that the frequency of disruption, 
and subsequent disk creation events increases with increasing stellar effective temperature, 
corresponding to younger post-main sequence ages.

For typical white dwarfs of log $g=8.0$, it takes around 0.07 Gyr to cool to an effective 
temperature of 20,000 K, about 0.2 Gyr to 15,000 K, and just over 0.6 Myr to achieve 10,000 
K \citep{ber95a,ber95b}.  The only known disk-bearing white dwarf which is likely to have a 
cooling age significantly older than several hundred Myr is G166-58; at 7400 K and assuming 
log $g=8.0$, its cooling age should be near 1.3 Gyr.  Notably, the properties of G166-58 are 
rather anomalous compared to the other known disk-bearing white dwarfs: its infrared excess 
comes up at 5 $\mu$m \citep{far08a}, while it has a modest calcium abundance and accretion 
rate.  One possibility is that planetary system remnants at white dwarfs tend to stabilize by 
roughly 1 Gyr.  Minor planet belts may become significantly depleted on these timescales or 
gravitational perturbations may subside on shorter timescales via dynamical rearrangement 
\citep{jur08,bot05,deb02}.

On the other hand, the lack of dust at the relatively warm, but highly polluted white dwarf
HS 2253$+$803 points to a disk lifetime shorter than the timescale for removal of accreted
metals.  This very metal-enriched and carbon-poor degenerate sits in the outlying region of 
Figure \ref{fig12} where it is virtually the only star in this abundance range without a disk.  
Therefore, it is likely this star had a dust disk which has been fully consumed within the 
$10^6$ yr diffusion timescale for this warm DBAZ.

\subsection{Metal Accretion Rates}

It has been previously argued that DAZ white dwarfs with the highest calcium abundances 
also have an infrared excess and, thus, a circumstellar disk \citep{kil06}.  This argument has 
been slightly recast \citep{jur07b,jur08} to suggest that those DAZ white dwarfs with the highest
metal accretion rates are the stars with infrared excess.  However, put into this proper context, 
both DAZ and DBZ white dwarfs should exhibit this correlation, if correct.  This connection 
between disk frequency and metal accretion rate is now re-evaluated for all 53 metal-polluted 
white dwarfs observed with the IRAC camera.

Assuming accretion-diffusion equilibrium (i.e. a steady state), mass accretion rates were 
calculated using Equation 2 of \citet{koe06} with the best available photospheric calcium 
abundance determinations, together with the solar calcium abundance relative to either 
hydrogen or helium, as appropriate.  Settling times for various metals are usually within a 
factor of 2 \citep{koe06,dup93}, and calcium is employed here because it is the best studied 
element.  Where available for DAZ stars, convective envelope masses and calcium diffusion 
timescales were taken from Table 3 of \citet{koe06}.  Otherwise log $g=8$ was assumed and 
diffusion timescales were taken directly from their Table 2, while convective envelope masses 
were interpolated using their Table 3 values for stars of similar effective temperature and surface 
gravity (log $g=8$ was also assumed for G166-58 for reasons discussed in \citealt{far08b}).  For 
DBZ stars, convective envelope masses were read from Figure 1 of \citet{paq86}, and calcium 
diffusion times were interpolated using their Table 2 values; all assuming $M=0.6$ $M_{\odot}$.  

\begin{figure}
\plotone{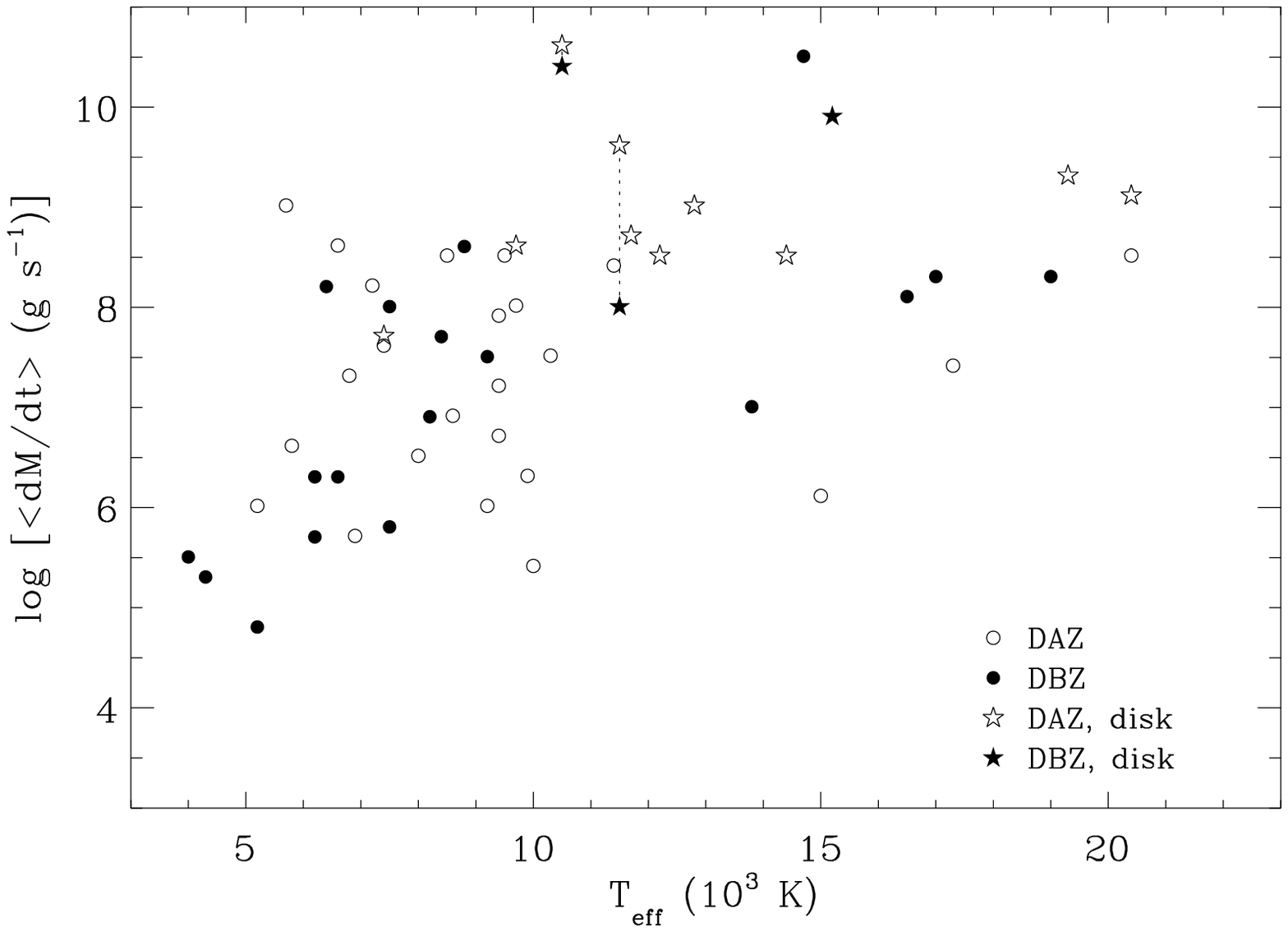}
\caption{Time-averaged dust accretion rates for all metal-contaminated white dwarfs studied 
by {\em Spitzer} IRAC.  Calculations were made using calcium diffusion times and convective 
envelope mass fractions from \citet{koe06} for DAZ stars and from \citet{paq86} for DBZ stars.   
In the plot, GD 16 and GD 362, both of spectral type DAZB (Table \ref{tbl4}), are plotted both as 
DAZ and as DBZ stars, their data points connected by dotted lines.  A typical uncertainty in the
accretion rate is 25\%.
\label{fig14}}
\end{figure}

\begin{figure}
\plotone{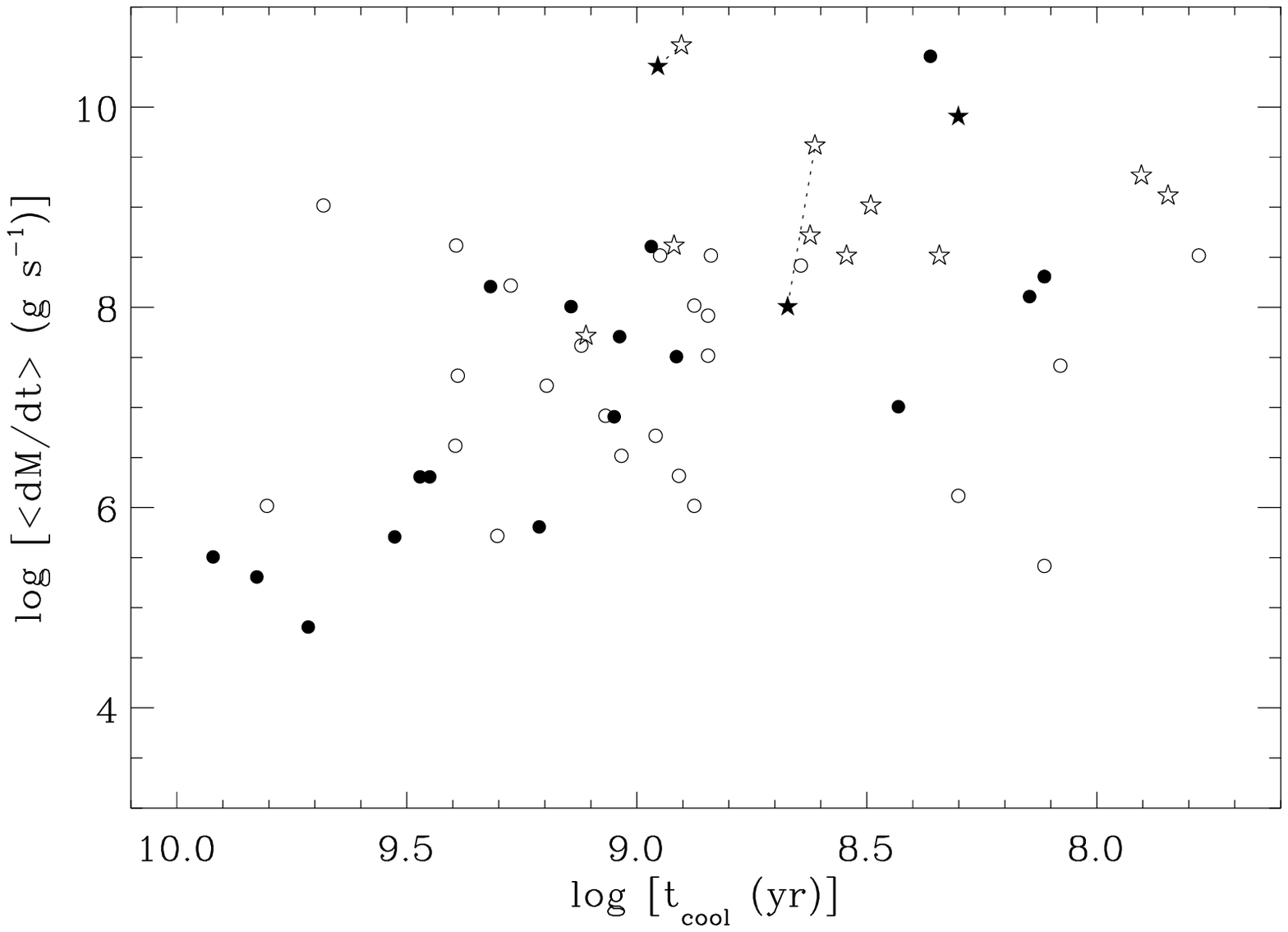}
\caption{Same as Figure \ref{fig14} but plotted versus cooling age.
\label{fig15}}
\end{figure}

In actuality, steady state accretion is likely for the warmer DAZ stars, but much less so for
the cooler DAZ and DBZ stars, which have relatively long metal dwell times.  Based on this
work and the results of \citet{kil08}, it is likely that disks at white dwarfs typically dissipate 
within $10^4-10^5$ yr, a timescale at which photospheric metals persist in the cooler DAZ 
and DBZ stars.  Hence, the calculated accretion rates listed in Table \ref{tbl4} should be 
considered {\em time-averaged} over a single diffusion timescale, and may not accurately
describe stars where detectable metals may dwell beyond $10^3$ yr.  An additional factor 
of $1/100$ was introduced to ``correct'' the \citet{koe06} Equation 2 rates to reflect that only 
accreted heavy elements are of interest, this factor being the typical dust-to-gas ratio in the 
interstellar medium.  Figure \ref{fig14} plots these accretion rates for all {\em Spitzer} observed 
stars versus effective temperature.  Figure \ref{fig15} plots the same metal accretion rates versus 
white dwarf cooling age.

The results show that the implied metal accretion rates are quite similar between the 
two atmospheric varieties of metal-rich white dwarf, strengthening the argument that 
their externally-polluted photospheres are caused by a common phenomenon; namely 
circumstellar material.  In fact, at accretion rates $dM/dt\ga$ $3\times10^8$ g s$^{-1}$, 
the fraction of metal-rich white dwarfs with circumstellar dust -- regardless of atmospheric 
composition -- is over 50\%.  If one associates accretion rate with circumstellar disk 
mass (a modest assumption), then this picture is consistent with white dwarf dust disks 
being strongly linked to tidal disruption events involving fairly large minor planets; while 
less massive disrupted asteroids give rise to more tenuous (possibly shorter-lived, possibly 
gaseous) disks, and modest accretion rates.

While the white dwarfs with an infrared excess are likely accreting from tidally-disrupted 
minor planets, the origin of the pollution in high accretion rate white dwarfs without obvious 
evidence for a disk is uncertain.  \citet{jur08} has proposed that if multiple, small asteroids 
are tidally-disrupted, their debris self-collides so the dust is efficiently destroyed, and matter 
then accretes onto the white dwarf from an undetected gaseous disk.  Given that the typical
timescale for collisions within a white dwarf dust disk is hours, with disk velocities at hundreds
of km s$^{-1}$, it is also conceivable that a single, modest-sized planetesimal could grind itself
down to gas-sized material via self-collisions \citep{far08b}.  Another possibility is that the 
dwell time of the photospheric metals in the white dwarf is longer than the characteristic 
disk dissipation time \citep{kil08}. 

\subsection{Circumstellar Versus Interstellar Accretion}

Two competing models to explain the metal-contaminated photospheres of white dwarfs 
involve interstellar accretion and circumstellar accretion \citep{sio90,alc86}.  The SEDs of 
the mid-infrared excess at GD 16 and PG 1457$-$086 are well explained by a circumstellar 
dust disk arising from a tidally disrupted asteroid or similar planetesimal, but are not consistent 
with interstellar accretion.  The same argument applies to the other metal-rich white dwarfs with 
mid-infrared excess; simply stated, the compact size and olivine composition of the dust are at 
odds with expectations for disks formed through interstellar accretion \citep{jur07a,jur07b,rea05}.

\begin{figure}
\plotone{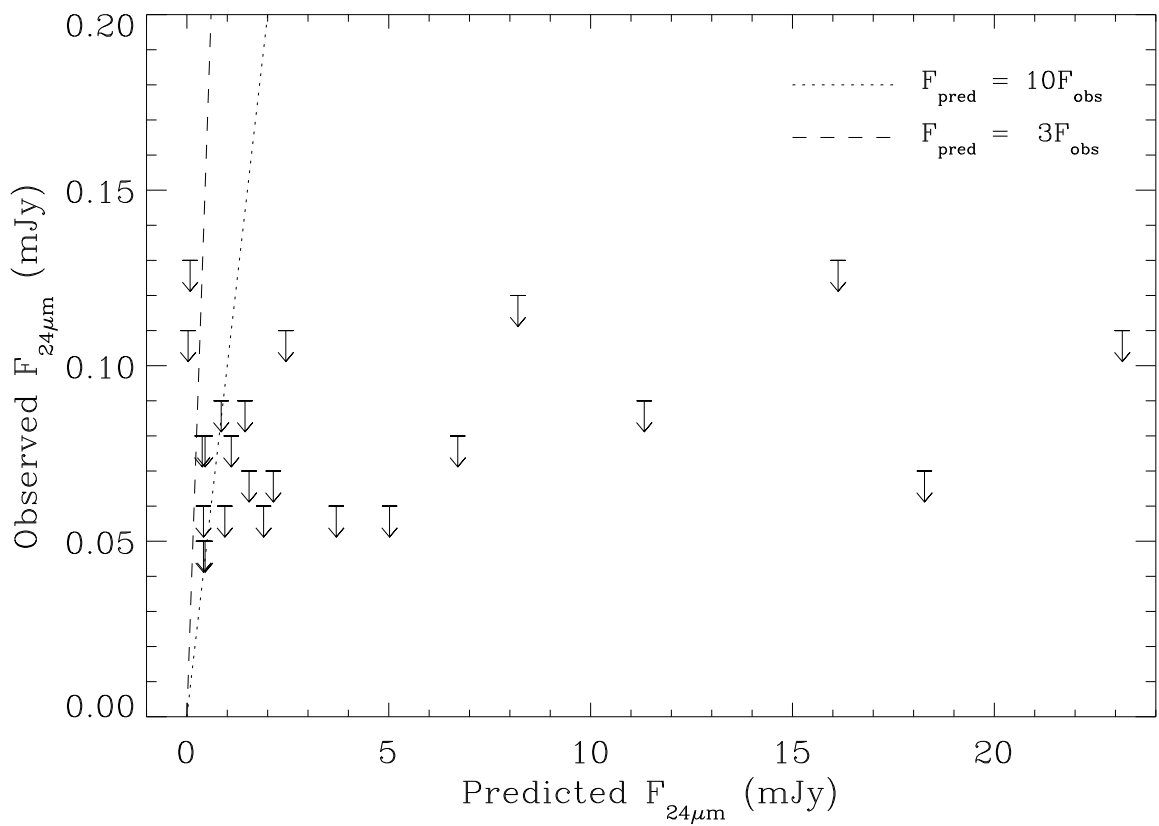}
\caption{Predicted infrared fluxes at 24 $\mu$m, assuming Bondi-Hoyle accretion of interstellar 
matter followed by Poynting-Robertson drag, plotted versus $3\sigma$ upper limits for 25 metal-rich 
white dwarfs observed with MIPS at this wavelength.  Stars with MIPS 24 $\mu$m detections that are 
consistent with warm circumstellar dust are not plotted, as well as those white dwarfs which suffered 
photometric contamination due to nearby sources.
\label{fig16}}
\end{figure}

With Equation 3 of \citet{jur07a} and following their method, predicted 24 $\mu$m 
fluxes were calculated for all 32 metal-rich white dwarfs observed with MIPS, assuming 
interstellar Bondi-Hoyle type grain accretion followed by Poynting-Robertson drag onto 
the star.  Figure \ref{fig16} displays these predicted infrared fluxes from interstellar accretion 
versus the MIPS 24 $\mu$m upper limits for 25 white dwarfs; stars with fluxes consistent with 
circumstellar disks were excluded, as were targets with contaminated photometry. The plot 
shows that the predicted fluxes are typically more than an order of magnitude greater than 
the observed $3\sigma$ upper limits.  With few possible exceptions, this simple interstellar 
accretion model cannot account for the observed infrared data.

Furthermore, there is now a growing number of DB white dwarfs with evidence for external 
pollution via carbon deficiency, relative to metals such as iron \citep{des08,zuc07,jur06,wol02}.  
They have accreted rocky material, independent of any evidence for or against circumstellar 
dust.  Table \ref{tbl5} lists the seven known helium- and metal-rich white dwarfs with low carbon 
abundances or upper limits.  Of these stars, only two of six observed by {\em Spitzer} have strong 
evidence in favor of circumstellar dust, GD 40 and GD 362.  This raises the possibility that the 
remaining dust-poor and metal-rich stars are accreting gaseous heavy elements.

Figure \ref{fig14} reveals that the DBZ HS 2253$+$803 has one of the highest implied 
(time-averaged) accretion rates among all polluted white dwarfs, yet has no dust disk as 
evidenced by the {\em Spitzer} photometry presented here.  This white dwarf is spectacularly 
carbon-poor relative to its metal content, and is highly likely to have accreted rocky material 
\citep{jur06}. In this particular case, its near pure helium nature gives it a likely diffusion 
timescale of $10^6$ yr, and hence a dissipated disk is quite possible; yet its atmospheric 
metal composition should nonetheless provide a measure of the extrasolar planetary 
material it has accreted.

The preponderance of the evidence firmly favors the interpretation that heavily 
metal-contaminated white dwarfs are currently accreting, or have in their recent history 
accreted, rocky planetesimal material in either dusty or gaseous form.  The origin of the
metals in white dwarfs at the low end of [Ca/H(e)] ratios and metal accretion rates, such 
as a number of DAZs in the survey of \citet{zuc03} remains unclear.

\subsection{Comparison with Main-Sequence Stars with Planetary Systems}

The circumstellar disks around white dwarfs likely arise from the tidal disruption 
of minor planets, with inferred metal accretion rates $dM/dt\ga$ $3\times10^{8}$ g 
s$^{-1}$.  In the solar system, the dust production rate in the zodiacal cloud is near $10^{6}$ 
g s$^{-1}$ \citep{fix02}.  Collisions among parent bodies result in dust production rates around 
main-sequence A-type stars in excess of $10^{10}$g s$^{-1}$ \citep{che06}.  Thus, the rate
estimated for the erosion of minor planets around white dwarfs is within the range inferred for 
main-sequence stars.  Because A-type stars evolve into white dwarfs, it appears there is an 
ample population of parent bodies among the main-sequence progenitors of white dwarfs to 
account for the observed pollutions.

The model in which an infrared excess around a white dwarf results from a tidally-disrupted 
minor planet requires that the orbit of an asteroid is perturbed substantially \citep{deb02}.  It 
is plausible that a Jupiter-mass planet is such a perturber.  \citet{cum08} find that 10.5\% of 
solar-type stars have gas giant planetary companions with periods between two and 2000 dy.
Clearly, many of these planets will be destroyed when the main-sequence star becomes a first 
ascent and then asymptotic giant \citep{far08a}, but half of these planets have periods longer 
than one year and may survive.  At present, the known fraction of main-sequence stars with 
massive planets that could persist beyond the post-main sequence evolution of their host is 
comparable to the estimated fraction of white dwarfs with an infrared excess.  The upper mass 
limits for self-luminous companions to white dwarfs, as found by this and previous {\em Spitzer} 
studies, are consistent with the scenario that asteroid orbits are perturbed by a typical gas giant 
planet \citep{far08a,far08b}.
 
\section{CONCLUSIONS}

{\em Spitzer} mid-infrared observations of metal-contaminated white dwarfs are extended to 
a larger sample, and to longer wavelengths, including numerous DBZ targets.  This study 
builds on previous work which all together indicate that warm dust orbits 21\% of all externally 
polluted white dwarfs observed by {\em Spitzer}.  Several patterns are revealed among the 
degenerates with and without warm circumstellar dust, which together lend support to the 
idea that tidally disrupted planetesimals are responsible for the heavy element abundances 
in many, if not most externally polluted white dwarfs. 

1.  Sufficient metal-rich targets now have been studied to estimate that between 1\% and 3\% of 
single white dwarfs with cooling ages less than around 0.5 Gyr possess an infrared excess that 
is likely the result of a tidally-disrupted asteroid.  Evidence is strong that white dwarfs can be used 
to study disrupted minor planets.

2.  As yet no evidence for cool, $T<400$ K dust exists from MIPS 24 $\mu$m observations 
of more than 30 metal-rich white dwarfs, including many with heavily polluted photospheres.
All stars with circumstellar disks detected at 24 $\mu$m have coexisting, strong $3-8$ $\mu$m 
IRAC excess fluxes.  The disks appear outwardly truncated; their mid-infrared spectral energy 
distributions are clearly decreasing at wavelengths beyond 8 $\mu$m.  These observed and 
modeled infrared excesses indicate rings of dust where the innermost grains typically exceed 
$T=1000$ K, and outer edges which lie within the Roche limit of the white dwarf, consistent 
with disks created via tidal disruption of rocky planetesimals. 

3.  Circumstellar disks at white dwarfs are vertically optically thick at wavelengths as long as 
20 $\mu$m.  Particles in an optically thin disk would not survive Poynting-Robertson drag for 
more than a few days to years.  Additionally, the warmest dust has been successfully modeled 
to lay within the radius at which blackbody grains in an optically thin cloud should sublimate 
rapidly.  With a single, notable exception (G166-58), the dusty circumstellar disks have inner 
edges which approach the sublimation region for silicate dust in an optically thick disk; precisely 
the behavior expected for a dust disk which is feeding heavy elements to the photosphere of its 
white dwarf host.

4. The majority of metal-contaminated white dwarfs do not have dust disks.  Circumstellar 
gas disks are a distinct possibility at dust-free, metal-rich stars with $dM/dt$ $\geq3\times
10^8$ g s$^{-1}$.  Fully accreted disks are a possibility for white dwarfs with metal diffusion 
timescales approaching $10^6$ yr.  It is possible that a critical mass and density must be 
reached to prevent the dust disk from rapid, collisional self-annihilation, and when this 
milestone is not reached, a gas disk results.  If correct, optical and ultraviolet spectroscopy 
of metal-rich white dwarfs are powerful tools with which to measure the bulk composition 
of extrasolar planetary material.

5.  Cooling age is correlated with the frequency of dusty disks at white dwarfs; for $T_{\rm eff}\la
20,000$ K, white dwarfs with younger cooling ages are more likely to be orbited by a dusty disk.  
G166-58 is by far the coolest white dwarf with a (rather anomalous) infrared excess, and the 
timescale for diffusion of metals out of the photosphere is relatively long for a DAZ white dwarf.  

{\em Spitzer} IRAC may soon bring a few more dust discoveries at polluted white dwarfs, 
but the statistics are unlikely to change significantly without a commensurate increase in the 
number of surveyed stars.  Such a program remains feasible with IRAC in the post-cryogenic 
phase, provided that $T\sim1000$ K dust is present in most white dwarf circumstellar disks.  
Unfortunately, objects with somewhat cooler dust emission such as G166-58 will evade detection 
until {\em JWST}, which should have photometric sensitivity similar to, or better than, {\em Spitzer} 
at all relevant mid-infrared wavelengths.  Observations of a sizable number of metal-contaminated 
white dwarfs at longer wavelengths, where cold planetesimal belt debris might be seen {\em in situ}, 
will have to await future facilities.  If belts of rocky planetary remnants persist around metal-polluted 
white dwarfs at tens of AU, where Poynting-Robertson drag cannot remove large particles within 
$10^9$ yr, then sensitive submillimeter observations may have a chance to directly detect them.

\acknowledgments

The authors thank the referee for constructive comments which improved the manuscript.  
J. Farihi thanks D. Koester and R. Napiwotzki for sharing aspects of their spectroscopic
dataset.  This work is based on observations made with the {\em Spitzer Space Telescope}, 
which is operated by the Jet Propulsion Laboratory, California Institute of Technology 
under a contract with NASA.  Support for this work was provided by NASA through an 
award issued by JPL/Caltech to UCLA.  This work has also been partly supported by 
the NSF.

{\em Facility:} \facility{Spitzer (IRAC,MIPS)}

\appendix

\section{IRAC CONSTRAINTS ON THE PRESENCE OF SUBSTELLAR COMPANIONS TO 
												METAL-RICH WHITE DWARFS}

An assessment of the presence of substellar companions to the metal-rich white dwarfs 
in this work and in \citet{jur07a} is made below via IRAC data.  Limits on unresolved 
substellar companions are placed at 23 metal-rich white dwarfs from IRAC 4.5 $\mu$m 
photometry and current models.  

\subsection{Limits for Unresolved Substellar Companions}

In addition to circumstellar dust, IRAC observations are sensitive to unresolved substellar 
companions at white dwarfs, via photometric excess \citep{far08a,mul07,deb07,han06,
far05b}.  Moreover, there exist a few metal-enriched white dwarfs which reside in close, 
detached, binaries with unevolved M dwarfs; in such systems, the source of the photospheric 
pollutants is thought to be the stellar wind of the main sequence companion \citep{zuc03}.  
Therefore, it is reasonable to ask if previously unseen, close, substellar companions might 
be responsible for the heavy elements in some contaminated white dwarfs.

Externally polluted white dwarfs observed with IRAC at 4.5 $\mu$m, but without obvious 
photometric excess due to circumstellar dust, and not previously analyzed for substellar 
companions, were selected from this study as well as from \citet{jur07a} and \citet{mul07}, 
yielding a total of 23 targets.  All data were analyzed independently as described in \S2.
Following the technique of \citet{far08b}, the absolute magnitude of a substellar companion 
with a putative $3\sigma$ photometric excess (above the observed flux) at 4.5 $\mu$m was 
determined using their Equation 2.  This magnitude was translated into a mass based on a 
calculated total (main sequence plus cooling) age for each white dwarf, utilizing appropriate 
substellar evolutionary models (I. Baraffe 2007, private communication; \citealt{bar03}). 

Table \ref{tbl6} lists the main sequence lifetime, $t_{\rm ms}$, cooling age, $t_{\rm cool}$,
distance from Earth, absolute 4.5 $\mu$m magnitude brightness limit and corresponding 
companion upper mass limit, for each metal-rich white dwarf.  Combined with previous results 
for 20 similar IRAC targets (16 from \citealt{far08a}; four from \citealt{deb07}), there are now 43 
cool contaminated white dwarfs observed with IRAC and analyzed for substellar companions; 
none show evidence for closely orbiting, potentially polluting secondaries down to an average 
mass of $19\pm8$ $M_{\rm J}$.  The hypothesis that unseen, brown dwarf companions are the 
source of heavy elements in some white dwarf atmospheres is essentially ruled out by these 
observations.

\subsection{Limits for Resolved Substellar Companions}

Again, following the methodology of \citet{far08a}, upper mass limits were established 
for spatially resolved (widely bound) T and sub-T dwarf companions for eight white dwarfs 
within $d\approx20$ pc.  These upper mass limits ($M_{\rm J}$, Table \ref{tbl7}) are 
considerably higher (average of 60 $M_{\rm J}$) than those for unresolved companions 
at the same stars, due to the criteria that they must be well-detected at all four IRAC 
channels \citep{far08a}.

\clearpage

\clearpage

\clearpage

\clearpage

\clearpage

\clearpage

\clearpage

\clearpage

\clearpage

\clearpage

\clearpage

\clearpage

\clearpage

\clearpage

\clearpage

\clearpage

\clearpage

\begin{deluxetable}{ccccccccc}
\tabletypesize{\footnotesize}
\tablecaption{White Dwarfs with Dust Disks\label{tbl1}}
\tablewidth{0pt}
\tablehead{
\colhead{WD}			&
\colhead{Name}		&
\colhead{SpT}			&
\colhead{$T_{\rm eff}$}	&
\colhead{$d$}			&
\colhead{$K$}			&
\colhead{Discovery}		&
\colhead{Discovery}		&
\colhead{Reference}\\

\colhead{}				&
\colhead{}				&
\colhead{}				&
\colhead{(K)}			&
\colhead{(pc)}			&
\colhead{(mag)}		&
\colhead{Year}			&
\colhead{Telescope}		&
\colhead{}}

\startdata

0146$+$187				&GD 16		&DAZB		&11500	&48		&15.3	&2008	&{\em Spitzer}		&1\\
0300$-$013				&GD 40		&DBZ		&15200	&74		&15.8	&2007	&{\em Spitzer}		&2\\
0408$-$041				&GD 56		&DAZ		&14400	&72		&15.1	&2006	&IRTF			&3\\
0842$+$231\tablenotemark{a}	&Ton 345 		&DBZ		&18600	&100	&15.9	&2008	&CFHT/Gemini		&4,5\\
1015$+$161				&PG			&DAZ		&19300	&91		&16.0	&2007	&{\em Spitzer}		&2\\
1041$+$092\tablenotemark{a}	&SDSS 1043 	&DAZ		&18300	&224	&\nodata	&2008	&{\em Spitzer}		&5\\
1116$+$026				&GD 133		&DAZ		&12200	&38		&14.6	&2007	&{\em Spitzer}		&2\\
1150$-$153				&EC			&DAZ		&12800	&76		&15.8	&2007	&IRTF			&6\\
1226$+$110\tablenotemark{a}	&SDSS 1228 	&DAZ		&22200	&142	&16.4	&2007	&{\em Spitzer}		&7\\
1455$+$298				&G166-58		&DAZ		&7400	&29		&14.7	&2008	&{\em Spitzer}		&8\\
1457$-$086      			&PG			&DAZ		&20400	&110	&16.0	&2008	&{\em Spitzer}		&1\\
1729$+$371				&GD 362		&DAZB		&10500	&57		&15.9	&2005	&IRTF/Gemini		&9,10\\
2115$-$560      			&LTT 8452	&DAZ		&9700 	&22		&14.0	&2007	&{\em Spitzer}		&11\\
2326$+$049      			&G29-38		&DAZ		&11700	&14		&12.7	&1987	&IRTF			&12\\

\enddata

\tablenotetext{a}{These stars with circumstellar gaseous metals are not analyzed in this paper.  
The WD numbers for SDSS 1228, SDSS 1043, and Ton 345 are unofficial, but correctly reflect the 
conventional use of epoch B1950 coordinates.}

\tablerefs{
(1) This work;
(2) \citealt{jur07a};
(3) \citealt{kil06};
(4) \citealt{mel08};
(5) C. Brinkworth 2008, private communication;
(6) \citealt{kil07};
(7) \citealt{bri08};
(8) \citealt{far08b};
(9) \citealt{bec05};
(10) \citealt{kil05};
(11) \citealt{von07};
(12) \citealt{zuc87b}}

\end{deluxetable}

\clearpage

\begin{deluxetable}{cccccc}
\tabletypesize{\scriptsize}
\tablecaption{Metal-Contaminated White Dwarf Targets\label{tbl2}}
\tablewidth{0pt}
\tablehead{
\colhead{WD}						&
\colhead{Name}					&
\colhead{SpT}						&
\colhead{$V$}						&
\colhead{[Ca/H(e)]\tablenotemark{a}}	&
\colhead{References}\\

\colhead{}							&
\colhead{}							&
\colhead{}							&
\colhead{(mag)}					&
\colhead{}							&
\colhead{}}

\startdata

0046$+$051				&vMa 2		&DZ		&12.39	&$-10.0$				&1\\
0146$+$187\tablenotemark{b}	&GD 16		&DAZB	&15.5	&$-5.8$\tablenotemark{c}	&2\\
						&			&		&		&$-8.7$\tablenotemark{d}	&2\\
0208$+$396				&G74-7		&DAZ	&14.51	&$-8.8$				&3\\
0738$-$172				&LHS 235		&DZA	&13.06	&$-10.9$				&1\\
1202$-$232				&EC			&DAZ	&12.79	&$-9.8$				&3\\
1257$+$278				&G149-28		&DAZ	&15.41	&$-8.1$				&3\\
1328$+$307				&G165-7		&DZ		&16.03	&$-8.1$				&1\\
1337$+$705				&G238-44		&DAZ	&12.77	&$-6.7$				&3\\
1457$-$086				&PG			&DAZ	&15.77	&$-6.1$				&4\\
1532$+$129\tablenotemark{b}	&G137-24		&DZ		&15.07	&$-8.4$				&5\\
1626$+$368				&G180-57		&DZA	&13.85	&$-8.8$				&1\\
1633$+$433				&G180-63		&DAZ	&14.84	&$-8.6$				&3\\
1653$+$385\tablenotemark{b}	&NLTT 43806	&DAZ	&15.9	&$-8.4$				&6\\
1705$+$030				&G139-13		&DZ		&15.20	&$-10.1$				&1\\
1858$+$393				&G205-52		&DAZ	&15.63	&$-7.8$				&3\\
2115$-$560				&LTT 8452	&DAZ	&14.28	&$-7.6$				&4\\
2215$+$388				&GD 401		&DZ		&16.20	&$-8.0$				&7\\
2251$-$070				&LHS 69		&DZ		&15.71	&$-10.5$				&1\\
2253$+$803				&HS			&DBAZ	&16.1	&$-6.1$				&8\\
2312$-$024				&G157-35		&DZ		&16.31	&$-10.6$				&1\\

\enddata

\tablenotetext{a}{A typical calcium abundance error is 0.1 dex.}

\tablenotetext{b}{The WD numbers for G137-24, GD 16, and NLTT 43806 are unofficial, 
but correctly reflect the conventional use of epoch B1950 coordinates.  Magnitudes with 
a single decimal precision are estimates and based on photographic data 
(e.g. \citealt{mcc06,mon03}).}

\tablenotetext{c}{[Ca/H]}	
\tablenotetext{d}{[Ca/He]}	

\tablerefs{
(1) \citealt{duf07};
(2) \citealt{koe05b};
(3) \citealt{zuc03};
(4) \citealt{koe05a};
(5) \citealt{kaw04};
(6) \citealt{kaw06};
(7) \citealt{dup93};
(8) \citealt{fri99}}

\end{deluxetable}

\clearpage

\begin{deluxetable}{cccccc}
\tabletypesize{\footnotesize}
\tablecaption{Mid-Infrared Fluxes and Upper Limits for White Dwarf Targets\label{tbl3}}
\tablewidth{0pt}
\tablehead{
\colhead{WD}					&
\colhead{$F_{3.6\mu{\rm m}}$}		&
\colhead{$F_{4.5\mu{\rm m}}$}		&
\colhead{$F_{5.7\mu{\rm m}}$}		&
\colhead{$F_{7.9\mu{\rm m}}$}		&
\colhead{$F_{24\mu{\rm m}}$ }\\

\colhead{}						&
\colhead{($\mu$Jy)}				&
\colhead{($\mu$Jy)}				&
\colhead{($\mu$Jy)}				&
\colhead{($\mu$Jy)}				&
\colhead{($\mu$Jy)}}

\startdata

0046$+$051\tablenotemark{a}	&$8040\pm400$	&$5360\pm270$	&$3680\pm190$	&$2080\pm110	$		&$110\pm$30\\
0108$+$277\tablenotemark{b}	&$357\pm56$		&$226\pm23$		&$140\pm19$		&$82\pm29$			&80\tablenotemark{c}\\
0146$+$187				&$486\pm24$		&$508\pm25$		&$473\pm24$		&$449\pm23$			&$120\pm40$\\
0208$+$396\tablenotemark{a}	&\nodata			&$723\pm36$		&\nodata			&$268\pm29$			&80\tablenotemark{c}\\	
0738$-$172				&$2930\pm150$	&$1810\pm90$		&$1350\pm100$	&$710\pm70$			&\nodata\\
1202$-$232\tablenotemark{a}	&\nodata			&$2010\pm100$	&\nodata			&$664\pm34$			&330\tablenotemark{c}\\
1257$+$278				&$290\pm15$		&$180\pm9$		&$128\pm7$		&$64\pm4$			&70\tablenotemark{c}\\
1328$+$307				&$225\pm11$		&$148\pm7$		&$96\pm6$		&$52\pm5$			&70\tablenotemark{c}\\
1334$+$039\tablenotemark{b}	&$2670\pm130$	&$1782\pm89$		&$1195\pm60$		&$698\pm35$			&70\tablenotemark{c}\\
1337$+$705\tablenotemark{a}	&\nodata			&$742\pm37$		&\nodata			&$241\pm14$			&$70\pm20$\\
1457$-$086				&$114\pm7$		&$76\pm6$		&$49\pm19$		&46\tablenotemark{c}	&110\tablenotemark{c}\\
1532$+$129				&$413\pm21$		&$275\pm14$		&$180\pm10$		&$112\pm7$			&60\tablenotemark{c}\\
1626$+$368				&$1033\pm52$		&$666\pm33$		&$441\pm22$		&$235\pm12$			&$60\pm20$\\
1633$+$433\tablenotemark{a}	&$912\pm46$		&$623\pm31$		&$389\pm20$		&$232\pm12$			&50\tablenotemark{c}\\
1653$+$385				&$230\pm12$		&$152\pm8$		&$93\pm6$		&$56\pm5$			&50\tablenotemark{c}\\
1705$+$030				&$507\pm25$		&$339\pm17$		&$217\pm12$		&$119\pm8$			&80\tablenotemark{c}\\
1858$+$393\tablenotemark{a}	&$201\pm10$		&$116\pm6$		&$73\pm6$		&$54\pm5$			&60\tablenotemark{c}\\
2115$-$560\tablenotemark{a}	&$1118\pm56$		&$1009\pm50$		&$991\pm50$		&$876\pm45$			&$230\pm30$\\
2215$+$388				&$110\pm6$		&$72\pm4$		&$43\pm4$		&$23\pm4$			&60\tablenotemark{c}\\
2251$-$070				&$1356\pm68$		&$929\pm46$		&$616\pm32$		&$406\pm23$			&130\tablenotemark{c}\\
2253$+$803				&$80\pm4$		&$50\pm3$		&$35\pm5$		&$19\pm3$			&70\tablenotemark{c}\\
2312$-$024				&$199\pm10$		&$133\pm7$		&$91\pm6$		&$43\pm6$			&90\tablenotemark{c}\\

\enddata

\tablecomments{Error calculations, including both photometric measurements and instrumental 
uncertainties are described in \S2.}

\tablenotetext{a}{IRAC fluxes for these objects have been previously reported \citep{far08a,far08b,deb07,mul07}.
The photometric measurements presented here were performed independently.}

\tablenotetext{b}{Not metal-polluted; see \S3.6}

\tablenotetext{c}{$3\sigma$ upper limit.}

\end{deluxetable}

\clearpage

\begin{deluxetable}{ccccccccc}
\tabletypesize{\footnotesize}
\tablecaption{Metal-Contaminated White Dwarfs Observed by {\em Spitzer}\label{tbl4}}
\tablewidth{0pt}
\tablehead{
\colhead{WD}				&
\colhead{Name}			&
\colhead{SpT}				&
\colhead{$M$}				&
\colhead{$T_{\rm eff}$}		&
\colhead{$t_{\rm cool}$}		&
\colhead{[H/He]}			&
\colhead{log $<dM/dt>$\tablenotemark{a}}		&
\colhead{References}\\

\colhead{}					&
\colhead{}					&
\colhead{}					&
\colhead{($M_{\odot}$)}		&
\colhead{(K)}				&
\colhead{(Gyr)}				&
\colhead{}					&
\colhead{(g s$^{-1}$)}		&
\colhead{}}

\startdata

0002$+$729				&GD 408		&DBZ	&0.59	&13800	&0.27				&$-6.0$				&7.0					&1,2\\
0032$-$175				&G266-135	&DAZ	&0.60	&9200	&0.75				&\nodata				&6.0					&3,4\\
0046$+$051				&vMa 2		&DZ		&0.69	&6200	&2.96				&$-3.2$				&6.3					&1,4,5\\
0146$+$187\tablenotemark{b}	&GD 16		&DAZB	&0.59	&11500	&0.47\tablenotemark{c}	&$-2.9$				&8.0\tablenotemark{c}	&5,6\\
						&			&		&		&		&0.41\tablenotemark{d}	&					&9.6\tablenotemark{d}	&\\	
0208$+$396				&G74-7		&DAZ	&0.66	&7200	&1.88				&\nodata				&8.2					&3,5,7\\
0235$+$064				&PG			&DAZ	&0.61	&15000	&0.20				&\nodata				&6.1					&4,7\\
0243$-$026				&LHS 1442	&DAZ	&0.70	&6800	&2.45				&\nodata				&7.3					&7,8\\
0245$+$541				&G174-14		&DAZ	&0.76	&5200	&6.37				&\nodata				&6.0					&3,7\\
0300$-$013\tablenotemark{b}	&GD 40		&DBZ	&0.59	&15300	&0.20				&$-6.0$				&9.9					&9,10\\
0322$-$019				&G77-50		&DZA	&0.60	&5200	&5.18				&\nodata				&4.8					&8,11\\
0408$-$041\tablenotemark{b}	&GD 56		&DAZ	&0.60	&14400	&0.22				&\nodata				&8.5					&8,10\\
0552$-$041				&G99-44		&DZ		&0.61	&4300	&6.70				&$-5.0$\tablenotemark{e}	&5.3					&2,9\\
0738$-$172				&LHS 235		&DZA	&0.62	&7600	&1.63				&$-3.4$				&5.8					&5,12\\
0843$+$358				&GD 95		&DZ		&0.58	&8200	&1.12				&$-5.5$\tablenotemark{e}	&6.9					&2,13\\
0846$+$346				&GD 96		&DAZ	&0.59	&7400	&1.32				&\nodata				&7.6					&3,4\\
1015$+$161\tablenotemark{b}	&PG			&DAZ	&0.61	&19300	&0.08				&\nodata				&9.3					&8,10\\
1102$-$183				&EC			&DAZ	&0.60	&8000	&1.08				&\nodata				&6.5					&3,4\\
1116$+$026\tablenotemark{b}	&GD 133		&DAZ	&0.59	&12200	&0.35				&\nodata				&8.5					&8,10\\
1124$-$293				&EC			&DAZ	&0.63	&9700	&0.75				&\nodata				&8.0					&4,8\\
1150$-$153\tablenotemark{b}	&EC			&DAZ	&0.60	&12800	&0.31				&\nodata				&9.0					&8,14\\
1202$-$232				&EC			&DAZ	&0.66	&8600	&1.17				&\nodata				&6.9					&2,3\\
1204$-$136				&EC			&DAZ	&0.61	&11400	&0.44				&\nodata				&8.4					&4,8\\
1208$+$576				&G197-47		&DAZ	&0.56	&5800	&2.48				&\nodata				&6.6					&3,4\\
1225$+$006				&HE			&DAZ	&0.66	&9400	&0.91				&\nodata				&6.7					&8,10\\
1257$+$278				&G149-28		&DAZ	&0.58	&8500	&0.89				&\nodata				&8.5					&3,5,7\\
1315$-$110				&HE			&DAZ	&0.86	&9400	&1.57				&\nodata				&7.2					&8,10\\
1328$+$307				&G165-7		&DZ		&0.57	&6400	&2.08				&$-3.0$				&8.2					&5,12\\
1337$+$705				&G238-44		&DAZ	&0.58	&20400	&0.06				&\nodata				&8.5					&2,3,5\\
1344$+$106				&G63-54		&DAZ	&0.65	&6900	&2.01				&\nodata				&5.7					&3,4\\
1407$+$425				&PG			&DAZ	&0.67	&9900	&0.81				&\nodata				&6.3					&3,4\\
1455$+$298\tablenotemark{b}	&G166-58		&DAZ	&0.58	&7400	&1.29				&\nodata				&7.7					&3,4\\
1457$-$086\tablenotemark{b}	&PG			&DAZ	&0.62	&20400	&0.07				&\nodata				&9.1					&5,8,10\\
1532$+$129				&G137-24		&DZ		&0.58	&7500	&1.39				&\nodata				&8.0					&5,15\\
1626$+$368				&G180-57		&DZA	&0.59	&8400	&1.09				&$-3.6$				&7.7					&5,12\\
1632$+$177				&PG			&DAZ	&0.58	&10000	&0.13				&\nodata				&5.4					&3,4\\
1633$+$433				&G180-63		&DAZ	&0.68	&6600	&2.47				&\nodata				&8.6					&3,4,5\\
1653$+$385				&NLTT 43806	&DAZ	&0.77	&5700	&4.80				&\nodata				&9.0					&5,16\\
1705$+$030				&G139-13		&DZ		&0.70	&6600	&2.82				&$-3.6$				&6.3					&5,12\\
1729$+$371\tablenotemark{b}	&GD 362		&DAZB	&0.73	&10500	&0.90\tablenotemark{c}	&$-1.1$				&10.4\tablenotemark{c}	&17,18\\
						&			&		&		&		&0.80\tablenotemark{d}	&					&10.6\tablenotemark{d}	&\\
1822$+$410				&GD 378		&DBAZ	&0.55	&17000	&0.13				&$-4.0$				&8.3					&1,2\\
1826$-$045				&G21-16		&DAZ	&0.73	&9400	&0.70				&\nodata				&7.9					&4,8\\
1858$+$393				&G205-52		&DAZ	&0.59	&9500	&0.69				&\nodata				&8.5					&3,4,5\\
2105$-$820				&LTT 8381	&DAZ	&0.60	&10300	&0.70				&\nodata				&7.5					&2,8\\
2115$-$560\tablenotemark{b}	&LTT 8452	&DAZ	&0.66	&9700	&0.83				&\nodata				&8.6					&2,5,8\\
2144$-$079				&G26-31		&DBZ	&0.55	&16500	&0.14				&\nodata				&8.1					&8,10\\
2149$+$021				&G93-48		&DAZ	&0.59	&17300	&0.12				&\nodata				&7.4					&2,8\\
2215$+$388				&GD 401		&DZ		&0.58	&8800	&0.93				&$-3.4$\tablenotemark{e}	&8.6					&5,13\\
2216$-$657				&LTT 8962	&DZ		&0.58	&9200	&0.82				&$-4.0$				&7.5					&1,2\\
2251$-$070				&LHS 69		&DZ		&0.58	&4000	&8.34				&$-6.0$\tablenotemark{e}	&5.5					&5,12\\
2253$+$803				&HS			&DBAZ	&0.59	&14700	&0.23				&$-5.5$				&10.5				&1,5\\
2312$-$024				&G157-35		&DZ		&0.69	&6200	&3.36				&$-4.9$				&5.7					&5,12\\
2326$+$049\tablenotemark{b}	&G29-38		&DAZ	&0.62	&11700	&0.42				&\nodata				&8.7					&8,4,19\\
2354$+$159				&PG			&DBZ	&0.55	&19000	&0.13				&\nodata				&8.3					&8,10\\

\enddata

\tablenotetext{a}{A typical error in the calculated accretion rate is 25\%, arising from a typical 0.1 dex error in the measured calcium abundance.}
\tablenotetext{b}{Circumstellar disk detected.}
\tablenotetext{c}{Calculated for a DBZ white dwarf.}
\tablenotetext{d}{Calculated for a DAZ white dwarf.}
\tablenotetext{e}{Upper limit.}

\tablerefs{
(1) \citealt{wol02};
(2) \citealt{mul07};
(3) \citealt{zuc03};
(4) \citealt{far08a};
(5) This work;
(6) \citealt{koe05b};
(7) \citealt{deb07};
(8) \citealt{koe05a};
(9) \citealt{vos07};
(10) \citealt{jur07a};
(11) \citealt{far08b};
(12) \citealt{duf07};
(13) \citealt{dup93};
(14) \citealt{jur09};
(15) \citealt{kaw04};
(16) \citealt{kaw06};
(17) \citealt{zuc07};
(18) \citealt{jur07b};
(19) \citealt{rea05}}
\end{deluxetable}

\clearpage

\begin{deluxetable}{ccccc}
\tabletypesize{\small}
\tablecaption{Carbon-Poor and Metal-Rich White Dwarfs\label{tbl5}}
\tablewidth{0pt}
\tablehead{
\colhead{WD}					&
\colhead{Name}				&
\colhead{[C/Fe]\tablenotemark{a}}	&
\colhead{Dust?}				&
\colhead{Reference}}

\startdata

0002$+$729	&GD 408		&$-0.6$				&N		&1\\
0100$-$068	&G270-124	&$-0.6$				&N		&1\\
0300$-$013	&GD 40		&$-1.0$				&Y		&2\\
0435$+$410	&GD 61		&$-1.2$\tablenotemark{b}	&\nodata	&\nodata\\
1626$+$368	&G180-57		&$-0.7$\tablenotemark{b}	&N		&3\\
1729$+$371	&GD 362		&$0.0$\tablenotemark{b}	&Y		&4\\
2253$+$803	&HS			&$-2.4$\tablenotemark{b}	&N		&3\\

\enddata

\tablenotetext{a}{Iron and carbon abundances and upper limits taken from \citet{des08,zuc07,wol02};
a typical error in this ratio is 0.3 dex.}

\tablenotetext{b}{Upper limit.}

\tablerefs{
(1) \citealt{mul07};
(2) \citealt{jur07a};
(3) This work;
(4) \citealt{jur07b}}

\end{deluxetable}

\clearpage

\begin{deluxetable}{ccccccc}
\tabletypesize{\small}
\tablecaption{Target Ages and Upper Mass Limits for Unresolved IRAC Companions\label{tbl6}}
\tablewidth{0pt}
\tablehead{
\colhead{WD}					&
\colhead{$t_{\rm ms}$}			&
\colhead{$t_{\rm cool}$}			&
\colhead{$t_{\rm total}$}			&
\colhead{$d$}					&
\colhead{$M_{4.5}$}				&
\colhead{Mass}\\

\colhead{}						&
\colhead{(Gyr)}					&
\colhead{(Gyr)}					&
\colhead{(Gyr)}					&
\colhead{(pc)}					&
\colhead{(mag)}				&
\colhead{($M_{\rm J}$)}}

\startdata

0002$+$729					&1.9		&0.3		&2.2		&35		&14.0	&18\\
0108$+$277\tablenotemark{a}		&0.2		&7.0		&7.2		&14		&15.3	&20\\
0552$-$041					&1.2		&6.7		&7.9		&6.5		&15.4	&18\\
0738$-$172					&1.1		&1.6		&2.7		&8.9		&14.8	&13\\
0843$+$358					&2.5		&1.1		&3.6		&24		&14.7	&17\\
1202$-$232					&0.6		&0.9		&1.5		&11		&14.2	&12\\
1225$+$006					&0.6		&0.9		&1.5		&30		&14.5	&11\\
1315$-$110					&0.1		&1.5		&1.6		&33		&14.7	&11\\
1328$+$307					&3.2		&2.1		&5.3		&30		&14.9	&18\\
1334$+$039\tablenotemark{a}		&6.3		&4.9		&11.2	&8.2		&15.0	&25\\
1337$+$705					&2.5		&0.1		&2.6		&25		&13.5	&25\\
1532$+$129					&2.5		&1.4		&3.9		&23		&14.8	&17\\
1626$+$368					&1.9		&1.0		&2.9		&16		&14.6	&15\\
1653$+$385					&0.2		&4.9		&5.1		&15		&16.3	&10\\
1705$+$030					&0.4		&2.7		&3.1		&18		&15.1	&12\\
1822$+$410					&6.3		&0.1		&6.4		&50		&13.3	&40\\
2105$-$820					&0.6		&0.7		&1.3		&18		&14.3	&11\\
2149$+$021					&1.9		&0.1		&2.0		&25		&13.5	&20\\
2144$-$079					&6.3		&0.1		&6.4		&69		&13.3	&40\\
2215$+$388					&2.5		&0.9		&3.4		&51		&14.4	&18\\
2216$-$657					&2.5		&0.8		&3.3		&24		&14.6	&17\\
2251$-$070					&2.5		&8.3		&10.7	&8.1		&15.7	&20\\
2253$+$803					&1.9		&0.2		&2.1		&85		&13.6	&20\\
2312$-$024					&0.4		&3.4		&3.8		&27		&15.2	&13\\
2354$+$159					&6.3		&0.1		&6.4		&105	&13.5	&40\\

\enddata

\tablenotetext{a}{Not metal-polluted; see \S3.6}

\tablecomments{Stellar parameters are taken from the literature (see Table \ref{tbl2} references).
For stars with no mass or surface gravity determination, log $g=8.0$ was assumed.}

\end{deluxetable}

\clearpage

\begin{deluxetable}{ccccccc}
\tabletypesize{\small}
\tablecaption{Target Ages and Upper Mass Limits for Resolved IRAC Companions\label{tbl7}}
\tablewidth{0pt}
\tablehead{
\colhead{WD}			&
\colhead{$t_{\rm ms}$}	&
\colhead{$t_{\rm cool}$}	&
\colhead{$t_{\rm total}$}	&
\colhead{$d$}			&
\colhead{$M_{7.9}$}		&
\colhead{Mass}\\

\colhead{}				&
\colhead{(Gyr)}			&
\colhead{(Gyr)}			&
\colhead{(Gyr)}			&
\colhead{(pc)}			&
\colhead{(mag)}		&
\colhead{($M_{\rm J}$)}}

\startdata

0108$+$277\tablenotemark{a}		&0.2		&7.0		&7.2		&14		&14.3	&60\\
0738$-$172					&1.1		&1.6		&2.7		&8.9		&14.0	&50\\
1334$+$039\tablenotemark{a}		&6.3		&4.9		&11.2	&8.2		&15.4	&50\\
1532$+$129					&2.5		&1.4		&3.9		&23		&13.2	&70\\
1626$+$368					&1.9		&1.0		&2.9		&16		&14.0	&50\\
1653$+$385					&0.2		&4.9		&5.1		&15		&14.1	&60\\
1705$+$030					&0.4		&2.7		&3.1		&18		&13.7	&60\\
2251$-$070					&2.5		&8.3		&10.7	&8.1		&14.7	&60\\

\enddata

\tablenotetext{a}{Not metal-polluted; see \S3.6}	

\tablecomments{The exposure time for 0738$-$172 was only 60 s versus 600 s for the primary
six targets, resulting in an overall sensitivity about $1/3$ that calculated in \citet{far08b}, while 
for 2251$-$070 the total integration time was 150 s and hence this target had an overall sensitivity 
about $1/2$ that of the primary target stars.}

\end{deluxetable}


\begin{thebibliography}{}

\bibitem[Alcock et al.(1986)Alcock, Fristrom, \& Siegelman]{alc86} Alcock, C., Fristrom, 
		C. C., \& Siegelman, R. 1986, \apj, 302, 462
	
\bibitem[Baraffe et al.(2003)]{bar03} Baraffe, I., Chabrier, G., Barman, T. S., Allard, 
		F., \& Hauschildt, P. H. 2003, \aap, 402, 701
		
\bibitem[Becklin et al.(2005)]{bec05} Becklin, E. E., Farihi, J., Jura, M., Song, I., Weinberger, 
		A. J., \& Zuckerman, B. 2005, \apj, 632, L119

\bibitem[Bergeron et al.(1995a)Bergeron, Saumon, \& Wesemael]{ber95a} Bergeron, P., 
		Saumon, D., \& Wesemael, F. 1995a, \apj, 443, 764

\bibitem[Bergeron et al.(1995b)Bergeron, Wesemael, \& Beauchamp]{ber95b} Bergeron, 
		P., Wesemael, F., \& Beauchamp, A. 1995b, \pasp, 107, 1047

\bibitem[Bottke et al.(2005)]{bot05} Bottke, W. F., Durda, D. D., Nesvorn\'y, D., Jedicke, 
		R., Morbidelli, A., Vokrouhlick\'y, D., \& Levison, Hal 2005. Icarus, 175, 111
	
\bibitem[Brinkworth et al.(2008)]{bri08} Brinkworth, C. S., G\"ansicke, B. T., Marsh, T. R.,
		Hoard, D. W., \& Tappert, C. 2009, \apj, in press

\bibitem[Burleigh et al.(2006)]{bur06} Burleigh, M. R., Hogan, E., Dobbie, P. D., 
		Napiwotzki, R., Maxted, P. F. L. 2006, MNRAS, 373, L55
		
\bibitem[Chen et al.(2006)]{che06} Chen, C. H., et al. 2006, \apjs, 166, 351	

\bibitem[Chiang \& Goldreich(1997)]{chi97} Chiang, E. I., \& Goldreich, P. 1997, \apj, 490, 368

\bibitem[Cumming et al. (2008)]{cum08} Cumming, A., Butler, R. P., Marcy, G. W., Vogt, S. S., 
		Wright, J. T., \& Fischer, D. A. 2008, \pasp, 120, 531

\bibitem[Dahn et al.(2002)]{dah02} Dahn, C. C., et al. 2002, \aj, 124, 1170
	
\bibitem[Debes \& Sigurdsson(2002)]{deb02} Debes, J. H., \& Sigurdsson, S. 2002, \apj, 572, 556
			
\bibitem[Debes et al.(2007)Debes, Sigurdsson, \& Hansen]{deb07}	Debes, J. H., 
		Sigurdsson, S. \& Hansen, B. 2007, \aj, 134, 1662

\bibitem[Desharnais et al.(2008)]{des08} Desharnais, S., Wesemael, F., Chayer, P., 
		Kruk, J. W., \& Saffer, R. A. 2008, \apj, 672, 540

\bibitem[Dufour et al.(2007)]{duf07} Dufour, P., et al. 2007, \apj, 663, 1291
	
\bibitem[Dupuis et al.(1993)Dupuis, Fontaine, \& Wesemael]{dup93} Dupuis, J., Fontaine, 
		G., \& Wesemael, F. 1993b, \apjs, 87, 345
	
\bibitem[Engelbracht et al.(2007)]{eng07} Engelbracht, C. W., et al. 2007, \pasp, 119, 994

\bibitem[Farihi(2005)]{far05c} Farihi, J. 2005, \aj, 129, 2382

\bibitem[Farihi et al.(2005a)Farihi, Becklin, \& Zuckerman]{far05a} Farihi, J., Becklin, 
		E. E., \& Zuckerman, B. 2005a, \apjs, 161, 394
		
\bibitem[Farihi et al.(2008a)Farihi, Becklin, \& Zuckerman]{far08a} Farihi, J., Becklin, E. E., 
		\& Zuckerman, B. 2008a, \apj, 681, 1470
	
\bibitem[Farihi et al.(2005b)Farihi, Zuckerman, \& Becklin]{far05b} Farihi, J., Zuckerman, 
		B., \& Becklin, E. E. 2005b, \aj, 130, 2237
				
\bibitem[Farihi et al.(2008b)Farihi, Zuckerman, \& Becklin]{far08b} Farihi, J., Zuckerman, B., 
		\& Becklin, E. E. 2008b, \apj, 674, 431

\bibitem[Fazio et al.(2004)]{faz04} Fazio, G. G., et al. 2004, \apjs, 154, 10

\bibitem[Fixsen \& Dwek(2002)]{fix02} Fixsen, D. J., \& Dwek, E. 2002, \apj, 578, 1009

\bibitem[Friedrich et al.(1999)]{fri99} Friedrich, S., Koester, D., Heber, U., Jeffery, C. S., 
		\& Reimers, D. 1999, \aap, 350, 865

\bibitem[Friedrich et al.(2007)]{fri07} Friedrich, S., Zinnecker, H., Correia, S., Brandner, W.,
		Burleigh, M. R., \& McCaughrean, M. 2007, Proceedings of the $15^{\rm th}$ 
		European Workshop on White Dwarfs, eds. M. R. Burlgeigh \& R. Napiwotzki
		(San Francisco: ASP), 343
		
\bibitem[G\"ansicke et al.(2008)]{gan08} G\"ansicke, B. T., Koester, D., Marsh, T. R., 
		Rebassa-Mansergas, A., \& Southworth J. 2008, \mnras, 391 L103
				
\bibitem[G\"ansicke et al.(2006)]{gan06} G\"ansicke, B. T., Marsh, T. R., Southworth, J., \& 
		Rebassa-Mansergas, A. 2006, Science, 314, 1908

\bibitem[G\"ansicke et al.(2007)G\"ansicke, Marsh, \& Southworth]{gan07} G\"ansicke, 
		B. T., Marsh, T. R., \& Southworth, J. 2007, \mnras, 380, L35
		
\bibitem[Giclas et al.(1965)Giclas, Burnham, \& Thomas]{gic65} Giclas, H. L., Burnham, 
		R. T., \& Thomas, N. G. 1965, Lowell Observatory Bulletin (Flagstaff: Lowell 
		Observatory), 6, 155	
	
\bibitem[Graham et al.(1990)]{gra90} Graham, J. R., Matthews, K., Neugebauer, G., \& 
		Soifer, B. T. 1990, \apj, 357, 216 

\bibitem[Greenstein(1984)]{gre84} Greenstein, J. L. 1984, \apj, 276, 602

\bibitem[Hansen et al.(2006)Hansen, Kulkarni, \& Wiktorowicz]{han06} Hansen, B. 
		M. S., Kulkarni, S., \& Wiktorowicz, S. 2006, \aj, 131, 1106

\bibitem[Hoard et al.(2007)]{hoa07} Hoard, D. W., Wachter, S., Sturch, L. K., Widhalm, 
		A. M., Weiler, K. P., Pretorius, M. L., Wellhouse, J. W., Gibiansky, M. 2007, 
		\aj, 134, 26
		
\bibitem[Holberg et al.(1997)]{hol97} Holberg, J. B., Barstow, M. A., \& Green, E. M. 
		1997, \apj, 474, L127

\bibitem[Holberg et al.(2002)Holberg, Oswalt, \& Sion]{hol02} Holberg, J. B., Oswalt, 
		T. D., \& Sion, E. M. 2002, \apj, 571, 512

\bibitem[Holberg et al.(2008)]{hol08} Holberg, J. B., Sion, E. M., Oswalt, T., McCook, G. P., 
		Foran, S., \& Subasavage, J. P. 2008, \aj, 135, 1225

\bibitem[Ida \& Lin(2008)]{ida08} Ida, S., \& Lin, D. N. C. 2008, \apj, 673, 487
								
\bibitem[Jura(2003)]{jur03} Jura, M. 2003, \apj, 584, L91

\bibitem[Jura(2006)]{jur06} Jura, M. 2006, \apj, 653, 613

\bibitem[Jura(2008)]{jur08} Jura, M. 2008, \aj, 135, 1785

\bibitem[Jura et al.(2007a)Jura, Farihi, \& Zuckerman]{jur07a} Jura, M., Farihi, J., \& 
		Zuckerman, B. 2007a, \apj, 663, 1285
		
\bibitem[Jura et al.(2009)]{jur09} Jura, M., Farihi, J., \& Zuckerman, B. 2009, AJ, in press
		
\bibitem[Jura et al.(2007b)]{jur07b} Jura, M., Farihi, J., Zuckerman, B., \& Becklin, E. E. 
		2007b, \aj, 133, 1927

\bibitem[Kawka \& Vennes(2006)]{kaw06} Kawka, A., \& Vennes, S. 2006, \apj, 643, 402

\bibitem[Kawka et al.(2004)Kawka, Vennes, \& Thorstensen]{kaw04} Kawka, A., Vennes, 
		S., \& Thorstensen, J R. 2004, \aj, 127, 1720

\bibitem[Kilic et al.(2008)]{kil08} Kilic, M., Farihi, J., Nitta, A., \& Leggett, S. K. 2008, \aj,
		136, 111
			
\bibitem[Kilic \& Redfield(2007)]{kil07} Kilic, M., \& Redfield, S. 2007, \apj, 660, 641 

\bibitem[Kilic et al.(2005)]{kil05} Kilic, M., von Hippel, T., Leggett, S. K., \& Winget, D. E. 
		2005, \apj, 632, L115
	
\bibitem[Kilic et al.(2006)]{kil06} Kilic, M., von Hippel, T., Leggett, S. K., \& Winget, D. E. 
		2006, \apj, 646, 474 
	
\bibitem[Koester \& Wilken(2006)]{koe06} Koester, D., \& Wilken, D. 2006, \aap, 453, 1051
	
\bibitem[Koester et al.(2005a)]{koe05a} Koester, D., Rollenhagen, K., Napiwotzki, R., Voss, 
		B., Christlieb, N., Homeier, D., \& Reimers, D. 2005a, \aap, 432, 1025 

\bibitem[Koester et al.(2005b)]{koe05b} Koester, D., Napiwotzki, R., Voss, B., Homeier, D., 
		\& Reimers, D. 2005b, \aap, 439, 317	

\bibitem[Lepine \& Shara(2005)]{lep05} Lepine, S., \& Shara, M. M. 2005, \aj, 129, 1483

\bibitem[Liebert(1977)]{lie77} Liebert, J. 1977, \aap, 56, 427
	
\bibitem[Liebert et al.(2005)Liebert, Bergeron, \& Holberg]{lie05} Liebert, J., Bergeron, P., 
		\& Holberg, J. B. 2005, \apjs, 156, 47

\bibitem[Marleau et al.(2004)]{mar04} Marleau, F. R., et al. 2004, \apjs, 154, 66
				
\bibitem[McCook \& Sion(2006)]{mcc06} McCook, G. P., \& Sion, E. M. 2006, Catalog of 
		Spectroscopically Identified White Dwarfs (Strasbourg: CDS)

\bibitem[McCook \& Sion(1999)]{mcc99} McCook, G. P., \& Sion, E. M. 1999, \apjs, 121, 1

\bibitem[Melis et al.(2008)Melis, Zuckerman, \& Albert]{mel08} Melis, C., Zuckerman, B.,
		Albert, L., Jura, M., \& Klein, B. 2008, \apj, submitted

\bibitem[Monet et al.(2003)]{mon03} Monet, D., et al. 2003, \aj, 125, 984
				
\bibitem[Mullally et al.(2007)]{mul07} Mullally, F., Kilic, M., Reach, W. T., Kuchner, M. J., 
		von Hippel, T., Burrows, A., \& Winget, D. E. 2007, \apjs, 171, 206

\bibitem[Napiwotzki et al.(2003)]{nap03} Napiwotzki, R., et al. 2003, Msngr, 112, 25

\bibitem[Paquette et al.(1986)]{paq86} Paquette, C., Pelletier, 	C., Fontaine, G., \& 
		Michaud, G. 1986, \apjs 61, 197

\bibitem[Perryman et al.(1997)]{per97} Perryman, M. A. C., et al. 1997, \aap, 323, L49

\bibitem[Probst(1983)]{pro83} Probst, R. 1983, \apjs, 53, 335

\bibitem[Reach et al.(2005)]{rea05} Reach, W. T., Kuchner, M. J., von Hippel, T., 
		Burrows, A., Mulally, F., Kilic, M., \& Winget, D. E. 2005, \apj, 635, L161

\bibitem[Rieke et al.(2004)]{rie04} Rieke, G., et al. 2004, \apjs, 154, 25

\bibitem[Salim \& Gould(2003)]{sal03} Salim, S., \& Gould, A. 2003, \apj, 582, 1011
	
\bibitem[Sion et al.(1990)]{sio90} Sion, E. M., Hammond, G. L., Wagner, R. M., Starrfield, 
		S. G., \& Liebert, J. 1990, \apj, 362, 691

\bibitem[Skrutskie et al.(2006)]{skr06} Skrutskie, M. F., et al. 2006, \aj, 131, 1163

\bibitem[von Hippel et al.(2007)]{von07} von Hippel, T., Kuchner, M. J., Kilic, M., Mullaly, 
		F., \& Reach, W. T. 2007, \apj, 662, 544

\bibitem[Voss et al.(2007)]{vos07} Voss, B., Koester, D., Napiwotzki, R., Christlieb, N., 
		\& Reimers, D. 2007, \aap, 470, 1079

\bibitem[Vrba et al.(2004)]{vrb04} Vrba, F., et al. 2004, \aj, 127, 2948

\bibitem[Wolff et al.(2002)Wolff, Koester, \& Liebert]{wol02} Wolff, B., Koester, D., \& Liebert, 
		J. 2002, \aap, 385, 995

\bibitem[Zuckerman (2001)]{zuc01} Zuckerman, B. 2001, \araa, 39, 549

\bibitem[Zuckerman \& Becklin(1987a)]{zuc87a} Zuckerman, B., \& Becklin, E. E. 1987a, 
		\apj, 319, 99
		
\bibitem[Zuckerman \& Becklin(1987b)]{zuc87b} Zuckerman, B., \& Becklin, E. E. 1987b,
		\nat, 330, 138
	
\bibitem[Zuckerman et al.(2007)]{zuc07} Zuckerman, B., Koester, D., Melis, C., Hansen, 
		B. M. S., \& Jura, M. 2007, \apj, 671, 872 
						
\bibitem[Zuckerman et al.(2003)]{zuc03} Zuckerman, B., Koester, D., Reid, I. N., \& 
		H\"unsch, M. 2003, \apj, 596, 477

\bibitem[Zuckerman \& Reid(1998)]{zuc98} Zuckerman, B., \& Reid, I. N. 1998, \apj, 505, 143
	
\end{thebibliography}
\end{document}